\documentclass[12pt]{article}

\usepackage{graphicx}
\usepackage{amsmath,amssymb,cite}
\usepackage{bm}
\usepackage{color}

\allowdisplaybreaks

\makeatletter
\@addtoreset{equation}{section}

\begin{document}
\baselineskip=0.6cm

\begin{flushright}
UT-KOMABA/08-4\\ 
KEK-TH-1151 \\ 
NSF-KITP-07-131 \\
February 2008
\end{flushright}
\vspace{0.3cm}
\begin{center}
\Large {\bf
A tunneling picture of dual giant Wilson loop}

\vspace{0.7cm}

\normalsize
 \vspace{0.4cm}
Akitsugu {\sc Miwa}$^a$\footnote{e-mail address:\ \ 
{\tt akitsugu@hep1.c.u-tokyo.ac.jp}}
,
Yoske {\sc Sumitomo}$^b$\footnote{e-mail address:\ \ 
{\tt sumitomo@post.kek.jp}}
and
Kentaroh {\sc  Yoshida}$^c$\footnote{e-mail address:\ \ 
{\tt kyoshida@kitp.ucsb.edu}}

\vspace{0.7cm}

$^a$ 
{\it Institute of Physics, University of Tokyo\\
Komaba, Meguro-ku, Tokyo 153-8902, Japan} \\

\vspace{0.4cm}

$^b$
{\it 
Department of Particle and Nuclear Physics, \\
The Graduate University for Advanced Studies, \\
Tsukuba, Ibaraki 305-0801, Japan.
}

\vspace{0.4cm}

$^c$
{\it Kavli Institute for Theoretical Physics, \\ 
University of California,  \\
Santa Barbara,  CA.\ 93106, USA. 
}

\vspace{1cm}
{\bf Abstract}
\end{center}

\noindent We further discuss a rotating dual giant Wilson loop (D3-brane) 
solution constructed in Lorentzian AdS by Drukker et al. The solution 
is shown to be composed of a dual giant Wilson loop and a dual giant graviton 
by minutely examining its shape. This observation suggests that 
the corresponding gauge-theory operator should be a 
$k$-th symmetric Wilson loop with the insertions of 
dual giant graviton operators.  
To support the correspondence, the classical action of the solution should 
be computed and compared with the gauge-theory result. 
For this purpose we first perform a Wick rotation to 
the Lorentzian solution by following the tunneling prescription 
and obtain Euclidean solutions corresponding to a circular 
or a straight-line Wilson loop. In Euclidean signature 
boundary terms can be properly considered in the standard manner 
and the classical action for the Euclidean solutions can be evaluated. 
The result indeed reproduces the expectation value of the $k$-th symmetric 
Wilson loop as well as the power-law behavior of the
correlation function of dual giant graviton operators.

\thispagestyle{empty}
\setcounter{page}{0}

\newpage
\section{Introduction}

One of the long-standing ideas in particle physics 
is to make a connection between a Wilson loop 
in gauge theory and a string-like object like in string theory. 
In the context of AdS/CFT correspondence \cite{AdS/CFT}, 
it is proposed that the expectation value of the fundamental 
Wilson loop is given by the ``area law'' of the fundamental string
world-sheet attached to the loop on the AdS boundary
\cite{Wilson1,Wilson2}. 
For straight lines and circular loops, the area
of the string world-sheet is shown to reproduce 
the expectation value of the Wilson loop calculated 
by summing up the planar ladder diagrams 
in a large 't Hooft coupling limit ($\lambda\equiv N g^2_{\rm YM}\to\infty$).

\medskip 

One may consider a multiply wrapped Wilson loop or a Wilson loop 
in higher-dimensional representation \cite{Drukker:2005kx}. 
It can carry a multiple winding number, say $k$, in terms of 
the fundamental representation. Hence a natural candidate 
for its counterpart is a state with string charge $k$\,. 
The multi-string state can be described as a spike D-brane solution 
with non-trivial electric flux describing the string charge \cite{Callan:1997kz}. 
It is now proposed that an anti-symmetric representation 
corresponds to an AdS$_2\times$S$^4$ D5-brane \cite{D5} 
called ``giant Wilson loop,'' and a symmetric representation 
to an AdS$_2\times$S$^2$ D3-brane \cite{Drukker:2005kx,GP1,GP2} 
called ``dual giant Wilson loop.''  
The names are analogy to (dual) 
giant gravitons \cite{GG,dGG1,dGG2}.

\medskip 

With the help of the string charge $k$\,, it is possible to consider a new
double-scaling limit, which is different from the usual 
large $N$ limit. In the case of $k$-th symmetric representation, 
$k$ and $N$ are taken to be
large while keeping $\kappa \equiv k \sqrt{\lambda} /4 N $ fixed. The
expectation value of the Wilson loop can be evaluated by using a Gaussian
matrix model and the result
completely agrees with the classical action 
of a D3-brane solution in the above limit \cite{Drukker:2005kx}.\footnote{
A symmetric Wilson loop cannot be distinguished from a
multiply wrapped one in the leading-order of approximation
at strong coupling \cite{Okuyama:2006jc,Hartnoll:2006is}.
See [14] for an argument on a
sub-leading contribution in the string side. 
The approach based on Gaussian matrix model 
was argued in \cite{Erickson:2000af,Drukker:2000rr}
and also in a recent work \cite{Pestun:2007rz}.
} 
Note that the classical action contains non-planar contributions 
in spite of large $N$, because large $k$ fundamental strings 
are bound on the D3-brane.

\medskip 

As another generalization,  
an R-charge $J$ may be introduced in analogy with \cite{BMN}.
A string solution rotating in S$^5$ 
has been constructed as a counterpart of a fundamental Wilson loop 
with local operator insertions $Z^J$ and its Hermitian conjugate 
\cite{Drukker:2006xg}. 
Here $Z$ is a complex scalar field in $\mathcal{N}$=4 SYM 
and related to a U(1) R-charge. 
Then an open spin chain description was discussed.

\medskip

Remember that the expectation values of Wilson loops are
usually discussed in Euclidean signature. Euclidean AdS is 
important also from the viewpoint of the bulk-boundary correspondence 
for local operators with R-charge. 
When in Lorentzian signature, the classical solutions 
that correspond to such operators are introduced 
at the center of AdS and do not reach the boundary. 
That is why a double Wick rotation has to be performed by following \cite{Dobashi:2002ar}. 
Then the bulk-boundary correspondence can be discussed by using the 
semi-classical bulk modes propagating along the ``tunneling trajectory,'' 
connecting the two points on the boundary. 

\medskip 

The tunneling method is also applicable to the fundamental 
Wilson loop with local operator 
insertions \cite{Yoneya:2006td,Miwa:2006vd}.\footnote{
A two-spin string around the tunneling trajectory is also 
discussed in \cite{Tsuji:2006zn},
and also a related work has been done 
in \cite{Zarembo:2002ph}. 
} 
The double Wick rotation for the Lorentzian solution \cite{Drukker:2006xg} 
leads to a Euclidean string solution which attaches to 
the Wilson loop on the boundary and propagates 
along the tunneling trajectory. Its classical action 
certainly reproduces the expectation value of the corresponding Wilson loop. 

\medskip 

In this paper we discuss a $k$-th symmetric Wilson loop carrying 
an R-charge. 
Then a D3-brane solution rotating in S$^5$ may be discussed 
as the string-theory counterpart (i.e., a rotating dual giant Wilson loop). 
In fact, a rotating D3-brane solution has already been constructed in
Lorentzian signature \cite{Drukker:2006zk}. 
Here we investigate the shape of the solution in detail. 
Then the solution is shown to be composed of a dual giant Wilson loop 
and a dual giant graviton \cite{dGG1,dGG2}, rather than a
rotating BPS particle. Thus this observation suggests 
that the dual gauge-theory operator should be a $k$-th symmetric Wilson loop 
with the insertions of dual giant graviton operators \cite{GGop,GGcorr}, 
rather than $Z^{J}$\,. 

\medskip 

Next we construct a Euclidean solution by applying the double Wick rotation 
for the Lorentzian solution. 
Then its classical action is evaluated by properly taking account of 
boundary terms. Although the computation is complicated 
the result is simple; The resulting action 
reproduces the expectation value of the $k$-th symmetric 
Wilson loop and also a two point function of the local operators 
with R-charge $J$ as it is expected. 

\medskip 

This paper is organized as follows: 
Section \ref{Tunneling_Review} is a brief review of the tunneling 
picture. Its new application to a dual giant graviton is also discussed. 
Section \ref{review} is also a review of the rotating D3-brane solution 
constructed in \cite{Drukker:2006zk}. We newly find the relation between the
solution and a dual giant graviton. This is the key observation 
to correctly identify the corresponding gauge-theory operator. 
In subsections \ref{tunnelingdGW} a double Wick rotation is performed to  
the Lorentzian solution by following the tunneling prescription. 
The resulting Euclidean solution attaches to a circle or a straight line 
on the boundary and propagates along the tunneling trajectory. 
In section \ref{Evaluation} the classical action of the Euclidean solution 
is evaluated. Then in section \ref{Wilson_Loop} we discuss the relation between the resulting action and the expectation value of the Wilson loop. 
Section \ref{Conclusion} is devoted to a conclusion and discussions.

\section{Tunneling picture of bulk-boundary correspondence}
\label{Tunneling_Review}

The bulk-boundary correspondence can be manifestly discussed 
in Euclidean formulation. For this purpose the Wick rotation 
should be performed. 
But note that we are interested in the case with 
an angular momentum, where a subtlety for the Wick rotation 
exists \cite{Dobashi:2002ar}. 
Then the tunneling prescription should be utilized. 
It would be available for later discussion 
to give a brief review of the tunneling prescription 
with the three examples: 1) a BPS particle (BMN case), 
2) a dual giant graviton, 3) a rotating string world-sheet.
Note that the cases 1) and 3) are just reviews of the preceding works, 
but the case 2) has not been discussed in the earlier literatures and 
this is the first attempt. 

\subsection{Tunneling trajectory of BPS particle}
\label{TNT}

We give a brief review of the tunneling prescription by taking 
a BPS particle rotating in S$^5$ with an angular momentum $J$\,. 
Here we assume that $J$ is much less than $N$\,.  

\medskip 

The AdS$_5\times$S$^5$ geometry in global coordinates is given by 
\begin{align}
 { ds^2 \over L^2} 
 & =  
 - \cosh^2 \rho dt^2 
 + d \rho^2 
 + \sinh^2 \rho 
 \big( 
 d \chi^2 
 + 
 \sin^2 \chi \big( d \varphi_1^2 + \sin^2 \varphi_1 d \varphi_2^2 \big)
 \big)
 + 
 d \theta^2 + \sin^2 \theta d \phi^2
 \,,
 \label{global_metric}\\[2mm]
 { C_4 \over L^4 }& =  
 \sinh^4 \rho\, \sin^2 \chi\, \sin \varphi_1\,
 dt 
 \wedge d \chi \wedge d \varphi_1 \wedge d \varphi_2\,,
 \label{C_4}
\end{align}
with a constant dilaton field. The S$^2$ metric in S$^5$ is 
explicitly written down, since we consider classical solutions which are 
localized with respect to
the remaining directions. Thus the ${\rm S}^5$ part of the RR potential 
$C_4$ is also irrelevant for the solutions.

\medskip

A null trajectory of a point particle rotating in S$^5$ is given by 
\begin{equation}
\rho = 0\,, \qquad \theta = {\pi \over 2}\,, \qquad \phi = t\,. 
\label{nullt}
\end{equation}
It is known that the string modes propagating along the trajectory 
correspond to local operators with large R-charge \cite{BMN}. 
But the trajectory does not reach the boundary and hence it is not 
available to discuss the correlation functions of the operators. 

\medskip

A solution for this issue was proposed in \cite{Dobashi:2002ar} and it  
is based on a semi-classical tunneling phenomenon. Hence the prescription 
is called ``tunneling picture.'' From now on let us see the tunneling picture. 
First the trajectory \eqref{nullt} should be recaptured with the 
Poincar\'e coordinates of the AdS$_5$ geometry,
in which the AdS$_5\times$S$^5$ metric becomes
\begin{equation}
{ ds^2 \over L^2} 
= {dZ^2 - (dX_0)^2 + (dX_1)^2 + (dX_2)^2 + (dX_3)^2 \over Z^2} 
+ d \theta^2 + \sin^2 \theta\, d \phi^2 \,. \label{met}
\end{equation}
Then the trajectory (\ref{nullt}) is mapped to the following trajectory:
\begin{equation}
Z = { \ell \over \cos t}\,, \quad  
X_0 = \ell \tan t\,, \quad 
X_1=X_2=X_3 = 0\,, \quad
\phi = t\,. \label{null}
\end{equation}
Here we have introduced a constant parameter $\ell$\,, 
which is related to the scale invariance of the metric (\ref{met}). 
The trajectory (\ref{null}) satisfies the null condition, 
$Z^{-2}(\dot Z^2 -\dot X_0^2) + \dot \phi^2 = 0$\,, 
and also the equations of motion $ \dot X_0 /Z^2
= 1/\ell $ and $\dot \phi =1$\,. 

\medskip 

Here let us see the motion of $Z$\,. The null condition and 
the equations of motion of $X_0$ and $\phi$ lead to  
\[
 \dot Z^2 + V(Z) = 0\,, \qquad  V(Z) \equiv  -Z^4/\ell^2 + Z^2\,.
\]
This equation suggests that the classical solution \eqref{null} does not
reach the boundary $Z=0$ because of the potential barrier coming from
$V(Z)$\,. Thus the trajectory that reaches the boundary is realized as
a trajectory that tunnels the potential barrier. 

\medskip

Such a tunneling trajectory was proposed in \cite{Dobashi:2002ar} 
via the Wick rotation with respect to the parameter $t$ as $t_{\rm E}=it$ 
as well as the target space time coordinate as $X_4 = i X_0$\,. 
We have to consider simultaneously whether 
the imaginary angular velocity 
or equivalently the Wick rotation with respect 
to the angular direction as $\phi_{\rm E} = i \phi$\,
and use the ansatz $\phi_{\rm E} = t_{\rm E}$\,. 
The resulting tunneling trajectory is given by 
\begin{equation}
Z= { \ell \over \cosh t_{\rm E}}\,,\qquad
X_1=X_2=X_3=0\,, \qquad 
X_4 = \ell \tanh t_{\rm E}\,. \label{tunneling} 
\end{equation}
This describes a semi-circle $Z^2+X_4^2=\ell^2$ in the $(Z,X_4)$ plane.

\medskip 

By considering modes propagating along the tunneling trajectory
\eqref{tunneling}, we can discuss a correlation function of the local
operators with R-charge.
The simplest example would be the two point
function of the BPS operators:
\begin{eqnarray}
\big\langle \,
{\rm Tr}_{\rm F} Z^J\!\big( \vec X_i \big) \,
{\rm Tr}_{\rm F} \overline Z^J\!\big( \vec X_f \big) \,
\big\rangle\,. 
\label{particle}
\end{eqnarray}
Here the complex scalar field $Z$ is defined as $\Phi_5 + i \Phi_6$.
The points $\vec X_{i,f}$ correspond to the two end points, 
$(X_1,X_2,X_3,X_4)=(0,0,0,\pm \ell)$, 
of the tunneling trajectory \eqref{tunneling}.  

\subsubsection*{Another derivation of tunneling null geodesic}
There is another derivation of the tunneling null geodesic (\ref{tunneling}). 
It should take the following steps: 
\begin{enumerate}
\item 
First let us consider the Euclidean AdS by performing 
the double Wick rotation: 
$t_{\rm E} = it$ and $\phi_{\rm E} = i \phi$\,.
\item Next we turn to the 
Euclidean Poincar\'e coordinates via
\begin{align}
& 
Z={ \ell \over f }, \quad 
X_1 = 
{ \ell \over f } 
\sinh \rho \sin \chi \sin \varphi_1 \cos \varphi_2\,, \quad
X_2 = 
{ \ell \over f } 
\sinh \rho \sin \chi \sin \varphi_1 \sin \varphi_2\,, 
\notag 
\\
&
X_3 
= 
{ \ell \over f } 
\sinh \rho 
( 
\alpha \sin \chi \cos \varphi_1 - \sqrt{1-\alpha^2} \cos \chi 
)\,, \quad
X_4 
= 
{ \ell \over f } \sinh t_{\rm E} \cosh \rho\,, \label{trans_2}\\
& 
f = 
\cosh t_{\rm E} \cosh \rho 
+ 
\sinh \rho 
( 
\sqrt{1-\alpha^2} \sin \chi \cos \varphi_1 + \alpha \cos \chi 
)\,, \qquad ( 0 \leq \alpha \leq 1 ). \notag  
\end{align}
\end{enumerate}
Here a constant parameter $\alpha$ is contained as well as $\ell$\,. 
It will be related to the shape of the Wilson loop later.
The transformation \eqref{trans_2}
can be decomposed into a series of simple coordinate transformations
as explained in Appendix \ref{transformation}. 

\medskip

After the double Wick rotation in the first step, 
the null trajectory \eqref{nullt} has been mapped to the trajectory 
described by $\rho=0$\,, $\theta=\pi/2$ and $\phi_{\rm E}=t_{\rm E}$\,.
Then by performing the transformation \eqref{trans_2}, 
it is mapped to the tunneling trajectory \eqref{tunneling}.

\medskip

In subsection \ref{tunnelingdGG} the above steps are applied to 
a dual giant graviton solution \cite{dGG1,dGG2}, which corresponds to a local operator carrying an R-charge of order $N$ or larger. 
In subsection \ref{ReviewDK_MY} we give a brief review of 
the Euclidean string solution of \cite{Yoneya:2006td,Miwa:2006vd}, 
which was constructed by applying the above steps
to the Lorentzian solution of \cite{Drukker:2006xg}.
In the next section we apply the above steps
to a dual giant Wilson loop (D3-brane) solution rotating in S$^5$ 
\cite{Drukker:2006zk}. 

\subsection{Dual giant graviton around tunneling trajectory} 
\label{tunnelingdGG}

From now on let us discuss the tunneling picture of 
a dual giant graviton solution. Here we assume that its angular momentum 
(R-charge) is the same order as $N$ or larger. 

\medskip 

The coordinates $t$\,, $\chi$\,, $\varphi_1$ and $\varphi_2$ are used 
as the world-volume coordinates of the dual giant graviton solution.  
Then the solution will be given by 
\begin{equation}
 \rho = \overline \rho:~{\rm const.}\,, \qquad
 \theta = {\pi \over 2}\,, \qquad
 \phi =t\,.  \label{dGG_sol}
\end{equation}
The double Wick rotation just changes the last equation of \eqref{dGG_sol}
to $\phi_{\rm E}=t_{\rm E}$.

\begin{figure}[tb]
\begin{center}
\begin{minipage}{5cm}
\includegraphics[width=5cm]{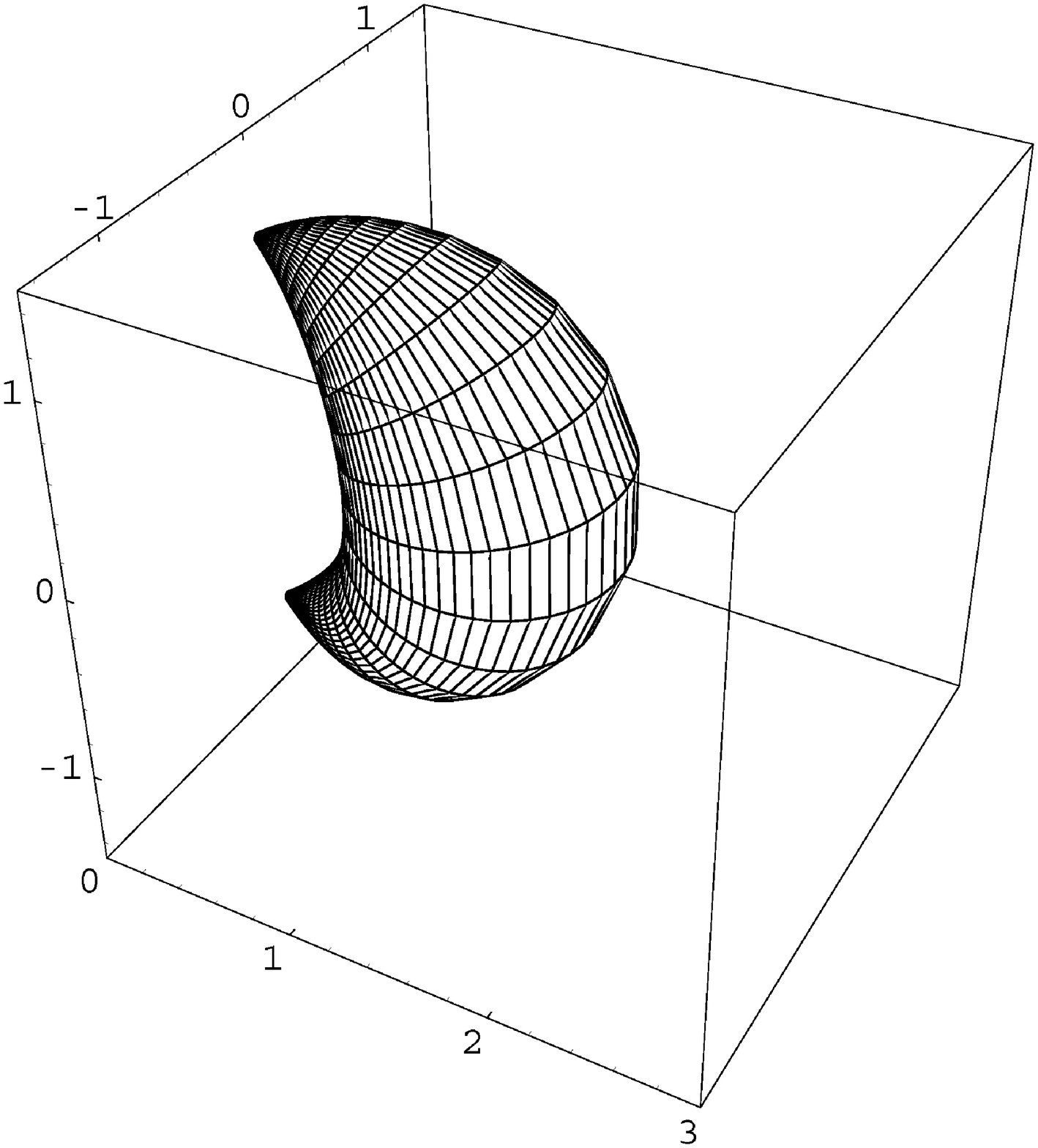}
\put(-108,10){\small{$Z/\ell$}}
\put(-170,76){\small{$X_4/\ell$}}
\put(-140,148){\small{$X_3/\ell$}} 
\\ \hspace{2cm} {\footnotesize (a)} 
\end{minipage} \hspace{3cm} 
\begin{minipage}{5cm}
\includegraphics[width=5cm]{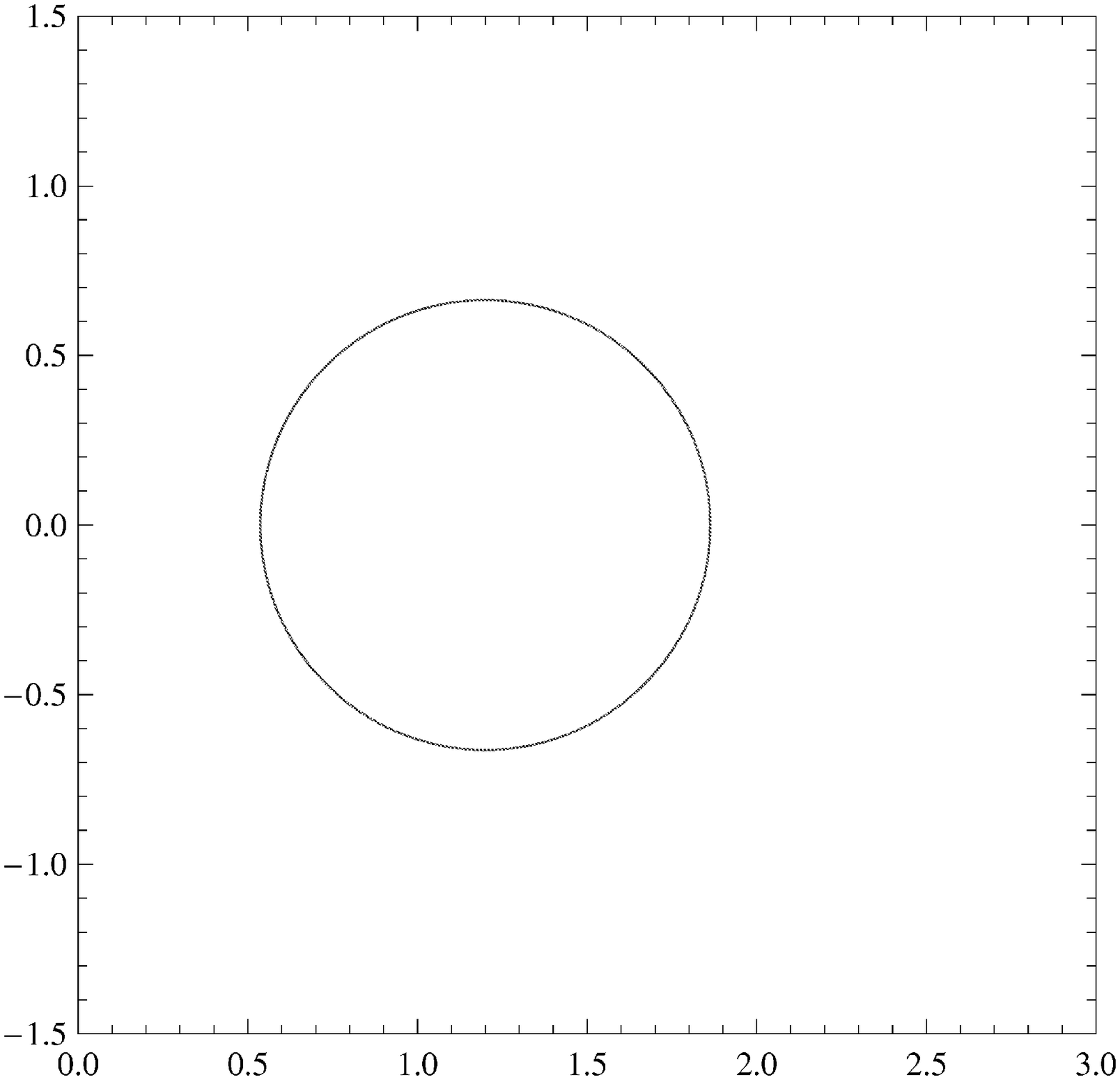}
\put(-78,-12){\small{$Z/\ell$}}
\put(-168,68){\small{$X_3/\ell$}} 
\vspace*{0.6cm}\\ \hspace*{2.3cm} {\footnotesize (b)}
\end{minipage}
\end{center}
\caption{\footnotesize Euclidean dual giant graviton. 
Its shape is depicted in (a). A time slice at $t=0$ of (a) is
 plotted in (b).} \label{dGG3D}
\end{figure}

\medskip 

Next we perform the transformation \eqref{trans_2}. 
For simplicity the case with $\alpha=0$ is discussed. 
Let us first concentrate on the slice of the solution on which 
the relation $\sin \varphi_1=0$ is
satisfied. This slice corresponds to the north and the south pole of the
${\rm S}^2$ spanned by $\varphi_1$ and
$\varphi_2$\,. Then the remaining directions of the world-volume are
two-dimensional. Indeed, for $\sin \varphi_1=0$\,, $X_1$
and $X_2$ vanish and the solution in terms of $(Z,X_3,X_4)$ 
is given by a two-dimensional surface depicted
in Fig.\,\ref{dGG3D}-(a). 

\medskip 

Figure \ref{dGG3D}-(b) is the cross section of Fig.\,\ref{dGG3D}-(a) 
at $t=0$\,, i.e., $(Z,X_3)$-plane at $X_4=0$\,.  The tunneling
trajectory penetrates the plane and the point is located 
at $( Z, X_3 ) = ( \ell , 0 )$\,. 
As far as $Z$, $X_3$ and $X_4$ are concerned, 
all other points on the D3-brane, i.e., the
region $0<\varphi_1<\pi$ are contained 
inside the surface of Fig.\,\ref{dGG3D}-(a).  
As for $X_1$ and $X_2$\,, points on the solution are in the region: 
$X_1^2 + X_2^2 \leq \ell^2 \sinh^2 \overline
\rho$\,. 

\medskip

The propagation of the dual giant graviton 
should correspond to a two point function of the 
dual giant graviton operators \cite{GGop,GGcorr}:
\begin{equation}
\big\langle \,
{\rm Tr}_{{\rm S}_J} Z \big( \vec X_i \big) \, 
{\rm Tr}_{{\rm S}_J} \overline Z \big( \vec X_f \big) \,
\big\rangle\,. \label{dGG_dGG}
\end{equation}
Here the trace is taken over the $J$-th symmetric representation. 
So far we have assumed that $J$ is the same order as $N$ or larger, 
but it may be possible to consider the limit $J \ll N$ in \eqref{dGG_dGG}. 
Then all of non-planar contributions 
in the dual giant graviton operator are negligible.
After all, \eqref{dGG_dGG} is reduced to \eqref{particle}. 
In the bulk gravity side a dual giant graviton 
shrinks in the same limit
and it should be regarded as a BPS particle.

\subsection{String world-sheet around tunneling trajectory}
\label{ReviewDK_MY}

Finally let us remember the tunneling picture of a rotating 
string world-sheet \cite{Yoneya:2006td,Miwa:2006vd}. 

\medskip 

We shall begin with the Lorentzian solution \cite{Drukker:2006xg}.
Taking $t$ and $\rho$ as world-sheet coordinates, 
it is given by 
\begin{equation}
\chi=0,\,\pi\,, \quad 
\varphi_1={\rm const.}\,, \quad 
\varphi_2={\rm const.}\,, \quad
\sin \theta = {1 \over \cosh \rho}\,, \quad
\phi = t\,. \label{DK_sol}
\end{equation}
Two patches, $\chi=0$ and $\chi=\pi$\,,  
are attached to straight lines on the boundary at $\rho=\infty$ 
and they are sewn together at $\rho=0$\,.
The solution carries an angular momentum from the infinite past 
$t=-\infty$ to the infinite future $t=\infty$\,. 

\medskip

As proposed in \cite{Drukker:2006xg}, a natural candidate 
for the dual gauge-theory operator 
would be the Wilson loop operator 
with local operator insertions\footnote{
See also the explanation in \cite{Drukker:2006zk}.}: 
\begin{equation}
W_{Z^J} \equiv {\rm Tr}_{\rm F}
{\cal P}
\Big[
Z^J(t=-\infty)
{\rm e}^{i\int_{-\infty}^\infty dt (A_t + \Phi_4)}
\overline Z^J (t=\infty) 
{\rm e}^{ i\int_\infty^{-\infty} dt (A_t + \Phi_4) }
\Big]\,. \label{ZZW}
\end{equation}
This operator contains two Wilson lines extending from 
$t=-\infty$ to $t=\infty$\,. Each of them corresponds to the line 
given by $(\rho, \chi) = (\infty,0)$ and $(\infty,\pi)$ 
on which the string world-sheet is attached. 
The local operators $Z^J$ and ${\overline Z}^J$
may be regarded as a
``creation'' and an ``annihilation'' operator of the R-charge,
respectively. The R-charge ``created'' by $Z^J$ at the infinite past is
carried by the rotating string to the infinite future and then it is
``annihilated'' by ${\overline Z}^J$\,. 

\medskip

Although it is interesting proposal, this Lorentzian picture 
is not available when calculating the expectation value of the operator 
via the classical string action.
This is because the angular momentum is carried from the 
infinite past to the infinite future, and it does not reach the boundary.
As a result, the operator insertions must be assumed 
at the infinite past and future. 
The situation is just the same as in 
the case of correlation functions of local operators with R-charge. 
In fact, by applying the steps introduced in subsection \ref{TNT},
we can construct a solution corresponding to a Wilson loop 
with the insertions of local operators in a finite region 
on the AdS boundary \cite{Yoneya:2006td,Miwa:2006vd}.

\medskip

\begin{figure}[tb]
\begin{center}
\begin{minipage}{3cm}
\includegraphics[width=3cm]{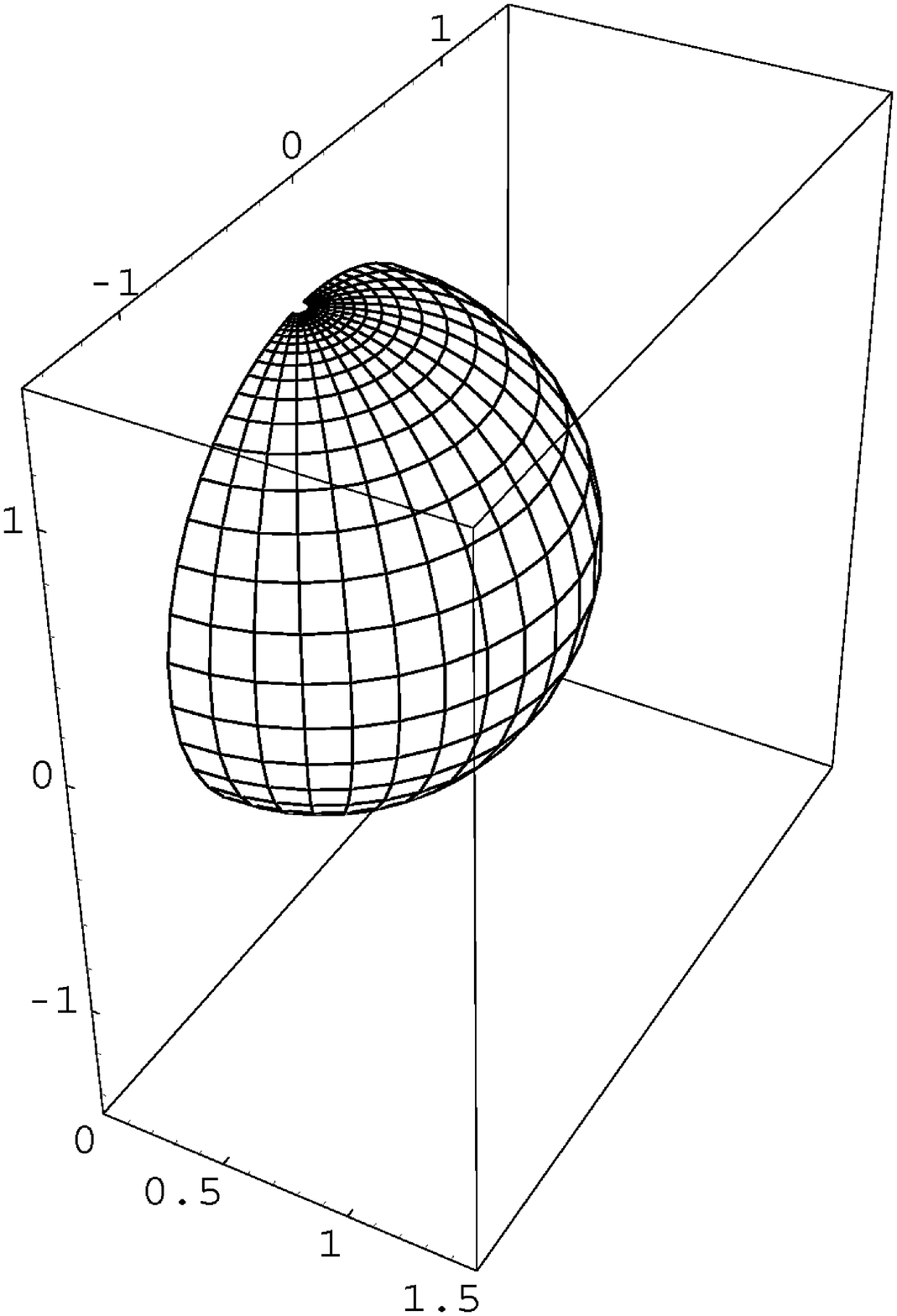}
\put(-75,0){\scriptsize $Z/\ell$}
\put(-77,118){\scriptsize $X_3/\ell$}
\put(-105,48){\scriptsize $X_4/\ell$}
\\ \hspace*{0.7cm} {\footnotesize (a)\quad $\alpha=0$} 
\end{minipage} \hspace{2cm}
\begin{minipage}{3cm}
\includegraphics[width=3cm]{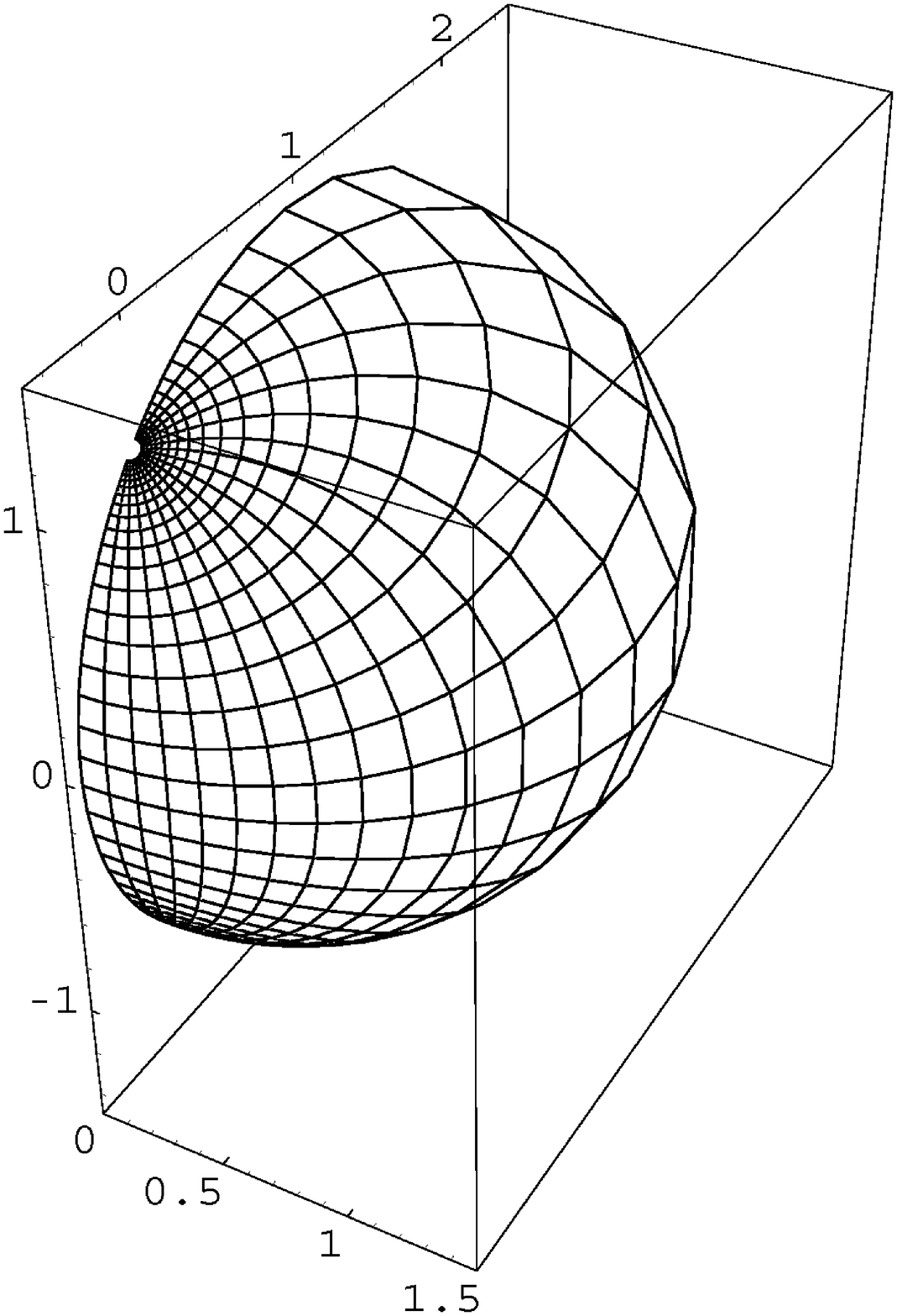}
\put(-75,0){\scriptsize $Z/\ell$}
\put(-77,118){\scriptsize $X_3/\ell$}
\put(-105,48){\scriptsize $X_4/\ell$}
\\ \hspace*{0.7cm} {\footnotesize (b)\quad $\alpha=0.7$} 
\end{minipage} \hspace{2cm}
\begin{minipage}{3cm}
\includegraphics[width=3cm]{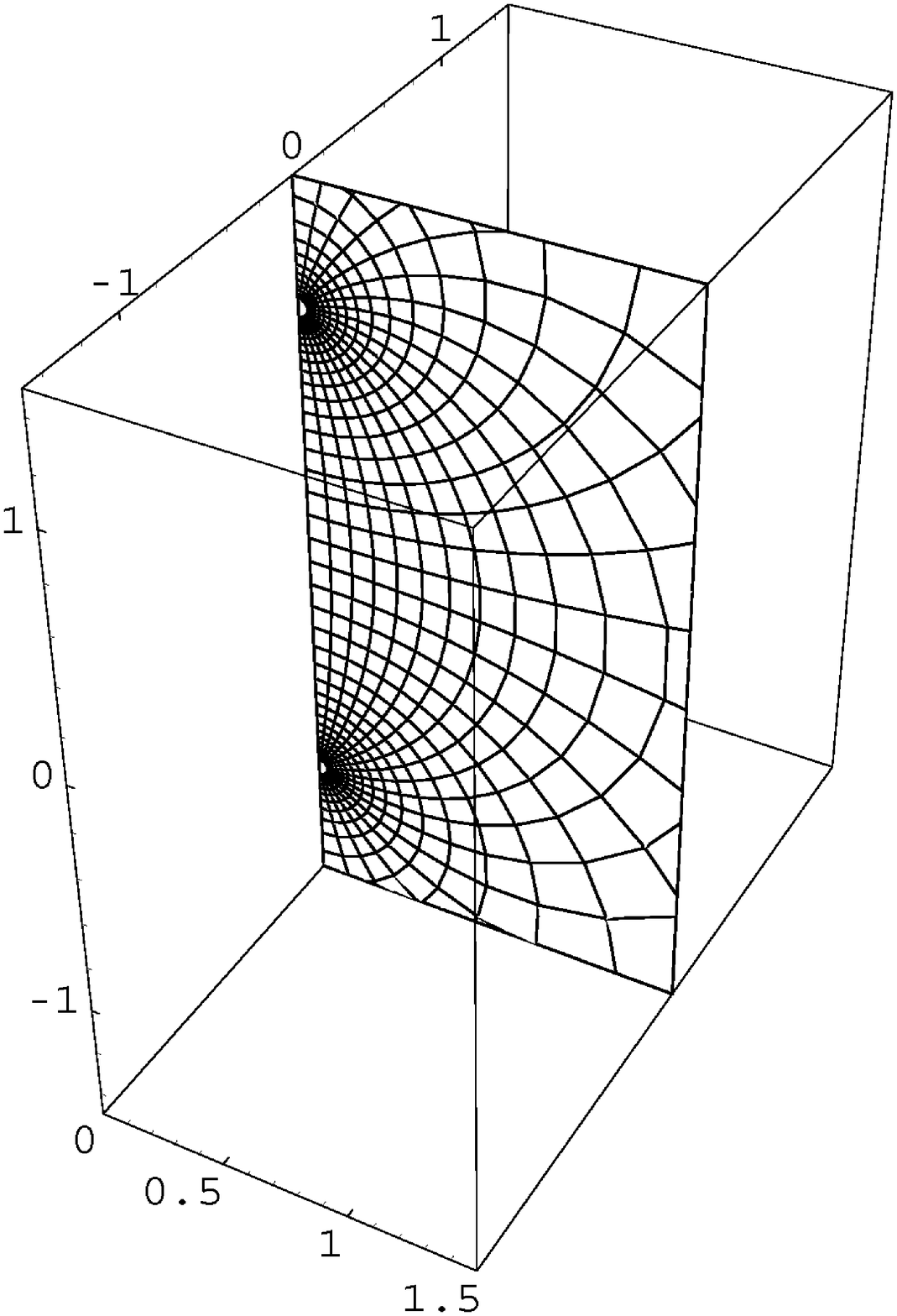}
\put(-75,0){\scriptsize $Z/\ell$}
\put(-77,118){\scriptsize $X_3/\ell$}
\put(-105,48){\scriptsize $X_4/\ell$}
\\ \hspace*{0.7cm} {\footnotesize (c)\quad $\alpha=1.0$}
\end{minipage}
\end{center}
\caption{\footnotesize Examples of string world-sheet attached to a circle or 
a straight line. Figures (a), (b) and (c) correspond 
to the case with $\alpha=0$, $0.7$ and $1.0$\,, respectively.
The world-sheet contains tunneling trajectory 
and carries an angular momentum along it.
Here $t$ and $\theta$ are used as parameters,
while $t$ and $\sigma \equiv {\rm arctanh}(\sin \theta)$ 
are used in \cite{Miwa:2006vd}.
}
\label{MY_fig}
\end{figure}

After performing the steps 1.~and 2.~in subsection \ref{TNT} 
to the solution \eqref{DK_sol}, 
the AdS$_5$ part of the resulting solution is given by 
\begin{align}
 & Z = { \ell \over \cosh t_{\rm E} \cosh \rho \pm \alpha \sinh \rho }\,,
 \quad 
 X_1 = X_2 = 0\,,  
 \notag 
 \\
 \quad 
 & X_3 = { \mp \ell \sqrt{1-\alpha^2} \sinh \rho \over \cosh t_{\rm E} \cosh \rho \pm \alpha \sinh \rho }\,,
 \quad
 X_4 = { \ell \sinh t_{\rm E} \cosh \rho \over \cosh t_{\rm E} \cosh \rho \pm \alpha \sinh \rho}\,. \label{MY_sol2}
\end{align}
This is the string solution constructed in \cite{Yoneya:2006td,Miwa:2006vd}.
Figures \ref{MY_fig}-(a), (b) and (c) depict the solutions
with $\alpha=0$, $0.7$ and $1.0$, respectively.

\medskip

By setting $\rho=0$ in 
\eqref{MY_sol2}, 
the solution contains the tunneling trajectory \eqref{tunneling}.
In fact, the solution carries angular momentum 
from the one end point $\vec X_i$ of 
the tunneling trajectory \eqref{tunneling} 
to the other end point $\vec X_f$\,.

\medskip

On the other hand, by taking large $\rho$ limit, 
the string world-sheet is attached to a circle ($\alpha \neq 1$) 
or a straight line ($\alpha=1$) on the AdS boundary $Z=0$\,.

\medskip 

It was shown in \cite{Miwa:2006vd} that 
the action of the string solution reproduces 
correct $\ell$- and $\alpha$-dependences of the 
expectation values of the following operator:
\begin{equation}
\bigg\langle 
{\rm Tr}_{\rm F}{\cal P}
\bigg[
\exp\bigg(
\oint_C ds 
\Big[ 
iA_\mu\big(\vec X(s)\big)\dot X^\mu (s)
+ 
\sqrt{\dot{\vec X}^2(s)}  \Phi^4\big(\vec X(s)\big) 
\Big]
\bigg)
Z^J \big(\vec X_i\big)  \overline Z^J \big(\vec X_f\big)
\bigg]
\bigg\rangle. 
\label{WZZ_MY}
\end{equation}
The shape of the loop $C$ is the same as that of the boundary 
of the solution \eqref{MY_sol2}.
Note that both of $\vec X_i$ and $\vec X_f$ 
are located on the loop $C$\,.

\medskip

Here the trace is taken over the fundamental representation. 
The aim of this paper is to extend the analysis to 
a $k$-th symmetric Wilson loop with local operator insertions.
In the next section we consider the tunneling picture of 
a rotating dual giant Wilson loop (D3-brane) solution.

\section{Tunneling picture of dual giant Wilson loop}
\label{review}

In this section we discuss a tunneling picture of dual giant Wilson loop. 
That is, the rotating string solution in
subsection \ref{ReviewDK_MY} is extended to a rotating D3-brane solution. 

\medskip 

We first reexamine the rotating D3-brane solution 
in Lorentzian AdS \cite{Drukker:2006zk}. It is a generalization of 
the string solution \eqref{DK_sol} to the D3-brane case. 
The shape of the solution leads us to observe that it is composed of 
a dual giant Wilson loop and a dual giant graviton. 

\medskip 

Then we have to perform the double Wick rotation 
and the coordinate transformation in subsection \ref{TNT}.
After that, the resulting solution is attached to a circle or straight line 
on the boundary of Euclidean Poincar\'e AdS and 
carrying an angular momentum from a point on the boundary to another. 

\subsection{Lorentzian solution and its properties}

Here we introduce a rotating dual giant Wilson loop (D3-brane) 
solution constructed in \cite{Drukker:2006zk}. 

\medskip 

Let us begin with the global coordinates \eqref{global_metric}. 
The coordinates $t$\,, $\rho$\,, $\varphi_1$ 
and $\varphi_2$ are regarded as the world-volume coordinates,
and the following ansatz is assumed for the region $0 \leq \chi \leq \pi/2$:
\begin{equation}
\chi = \chi(\rho)\,,  \quad 
\theta = \theta(\rho)\,,  \quad
\phi=t\,,  \quad
F_{t\rho} = \frac{L^2}{2\pi \alpha'}F(\rho)\,.
\label{ansatz}
\end{equation}
Here $ F_{t \rho} $ is an electric flux induced by smeared string
charges. Under this ansatz, the Dirac-Born-Infeld (DBI) action and the
Wess-Zumino (WZ) term for the region $0 \leq \chi \leq \pi/2$ are simplified as
\begin{align}
S_{\rm DBI}
&= 
-{2N \over \pi}
\int\! dt d \rho\,
\sinh^2 \rho\, \sin^2 \chi\,
\sqrt{
(\cosh^2 \rho - \sin^2 \theta)
(1 + \sinh^2 \rho\, {\chi'}^2 + {\theta'}^2) - F^2
}\,, \label{S_DBI} \\
S_{\rm WZ}
&= - {2N \over \pi} \int \! dt d\rho\,
\sinh^4 \rho\, \sin^2 \chi\, \chi'\,, \label{S_WZ}
\end{align}
where we have used the definition of D3-brane tension
\[
 T_{\rm D3} \equiv \frac{1}{(2\pi)^3l_s^4g_s} = \frac{N}{2\pi^2L^4}\,. 
\]
It is still difficult to find a classical solution even after assuming
the ansatz. A sensible way is to require the solution to
preserve some supersymmetries. Then it is possible to find a solution by
solving BPS equations rather than complicated equations of motion. In
fact, the solution concerned here has been derived by requiring a
quarter BPS condition \cite{Drukker:2006zk}.

\medskip 

The solution of \cite{Drukker:2006zk} is given by
\begin{align}
& \sin \chi(\rho) = { C_2 \coth \rho \over \sqrt{\cosh^2 \rho - C_1^2
}}\,, \qquad \sin \theta(\rho) = {C_1 \over \cosh \rho}\,,
\label{D3_chi_theta} \quad \\ 
& F(\rho) = - { \cosh^4 \rho - C_1^2 \over
\cosh^2 \rho \sqrt{\cosh^2 \rho - C_1^2 -C_2^2 \coth^2 \rho}}\,.
\label{D3_F}
\end{align}
For the region $\pi/2 \leq \chi \leq \pi$ the following replacement is
necessary: $\chi(\rho) \to \pi - \chi(\rho)$ and $F(\rho) \to
-F(\rho)$\,. 

\medskip

The two constant parameters $C_1$ and $C_2$ are related to 
two conserved charges, an angular momentum $J$ and a string charge $k$. 
The parameter $C_2$ is related to $k$ through 
\begin{equation}
C_2 = {k \sqrt{\lambda} \over 4 N} \equiv \kappa\,.
\end{equation}
Thus $C_2$ is nothing but $\kappa$ in the notation of
\cite{Drukker:2005kx}.
On the other hand, $J$ is given by 
\begin{equation}
J=2\times{2N \over \pi} C_2 C_1^2
\int \! d \rho\, 
\frac{
(\cosh^2 \rho - C_1^2)^2 + C_2^2 \cosh^4 \rho}
{(\cosh^2 \rho - C_1^2)^2 \cosh^2 \rho
\sqrt{\cosh^2 \rho - C_1^2 - C_2^2 \coth^2 \rho}
}\,. \label{J}
\end{equation}
The overall factor $2$ appears taking into account of the two patches.

\begin{figure}[tb]
\begin{center}
\begin{minipage}{5cm}
\begin{center}
\includegraphics[width=5cm]{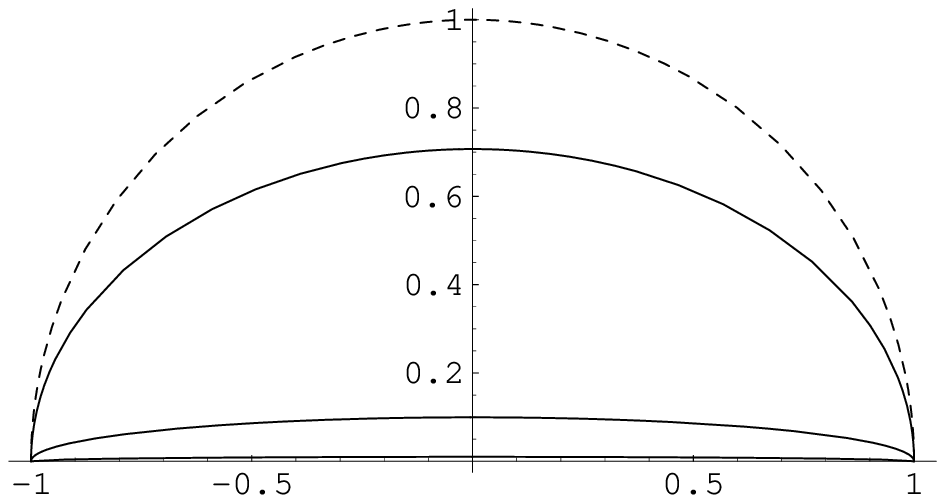} 
\put(-20,-3){{\tiny$\tanh \rho \cos \chi$}}
\put(-70,73){{\tiny$\tanh \rho \sin \chi$}}
\\
{\footnotesize (a)\quad $C_1=0$} 
\end{center}
\end{minipage} \hspace{1.5cm}
\begin{minipage}{5cm}
\begin{center}
\includegraphics[width=5cm]{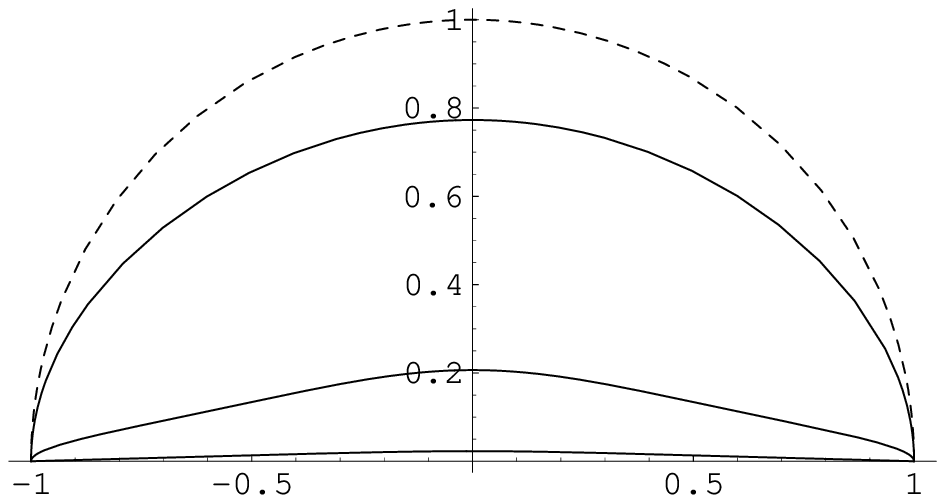} 
\put(-20,-3){{\tiny$\tanh \rho \cos \chi$}}
\put(-70,73){{\tiny$\tanh \rho \sin \chi$}}
\\
{\footnotesize (b)\quad $C_1=0.9$} 
\end{center}
\end{minipage}\\[3mm]
\begin{minipage}{5cm}
\begin{center}
\includegraphics[width=5cm]{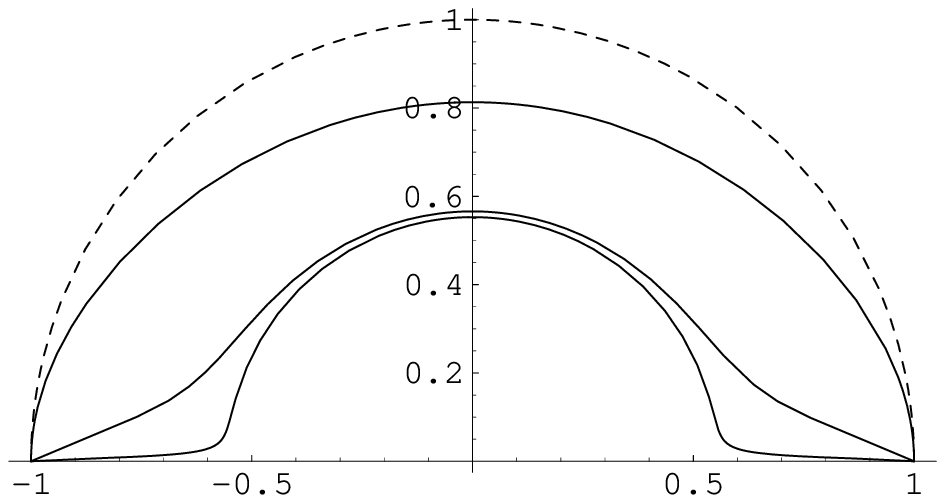} 
\put(-20,-3){{\tiny$\tanh \rho \cos \chi$}}
\put(-70,73){{\tiny$\tanh \rho \sin \chi$}}
\\
{\footnotesize (c)\quad $C_1=1.2$} 
\end{center}
\end{minipage} \hspace{1.5cm}
\begin{minipage}{5cm}
\begin{center}
\includegraphics[width=3cm]{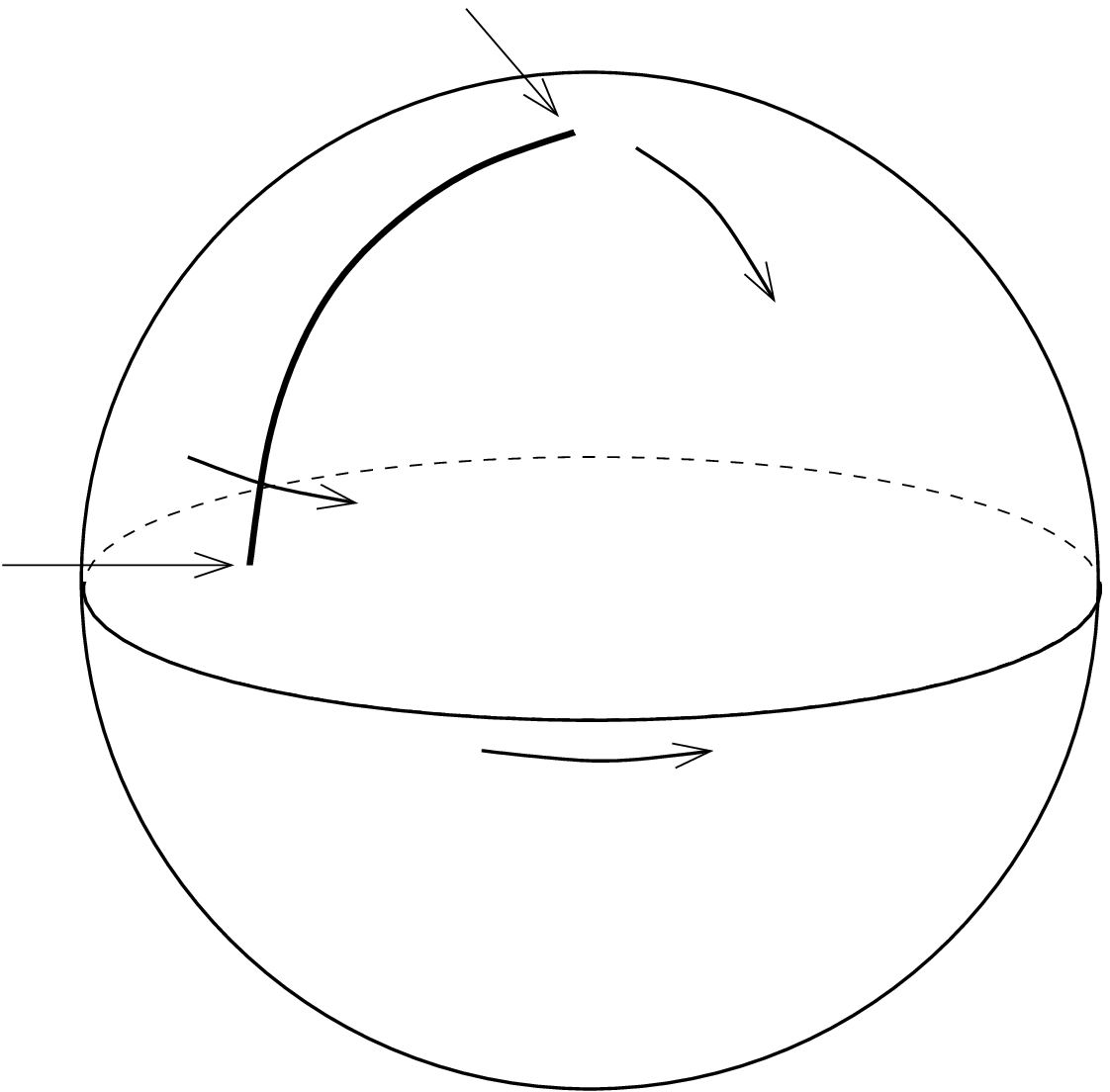} 
\put(-45,14){$\phi$}
\put(-30,65){$\theta$}
\put(-70,82){{\tiny $\rho=\infty$}}
\put(-115,38){{\tiny $\rho=\rho_{\rm min}$}}
\\
{\footnotesize (d) \quad S$^5$ part}
\end{center}
\end{minipage}
\end{center}
\caption{\footnotesize The D3-brane solution in the global coordinates. For various values of $C_1$ and $C_2$ it is numerically plotted.}
\label{dGW_global}
\end{figure}

\medskip 

Each D3-brane is attached to the AdS boundary $\rho=\infty$ at $\chi=0$
and $\pi$\,. 
Two patches with $0 \leq \chi \leq \pi/2$ and $\pi/2 \leq \chi \leq \pi$ are sewn
together smoothly at $\chi=\pi/2$\,. 
The radial coordinate $\rho$ takes the
minimal value, $\rho_{\rm min}$\,, at $\chi=\pi/2$\,.\footnote{
For the solution with $\chi(\rho)=0$\,, i.e., 
the solution with $C_1 \leq 1$ and $C_2 \to 0$\,, 
we define $\rho_{\rm min}=0$\,.}

\medskip

Some typical solutions are numerically plotted in Fig.\,\ref{dGW_global}\,. 
Figures \ref{dGW_global}-(a), (b) and (c) show 
the $(\rho,\chi)$ plane with an arbitrary $(t, \varphi_1, \varphi_2 )$ 
for $C_1=0$\,, $0.9$ and $1.2$, respectively. 
In particular, Fig.\,\ref{dGW_global}-(a) corresponds to the 
non-rotating Drukker-Fiol solution \cite{Drukker:2005kx}.
The radial and the angular coordinates of these figures are 
taken to be $\tanh \rho$ and $\chi$\,, respectively. 
Each broken line corresponds to the boundary 
of the ${\rm AdS}_5$~($\rho=\infty$)\, 
and three solid lines in each of the figures show the solutions with
$C_2 = 1.0$\,, $0.1$ and $0.01$ from the top down.

\medskip 

Figure \ref{dGW_global}-(d) describes a typical configuration 
$\theta = \theta(\rho)$ of the solution on 
${\rm S}^5$\, for a fixed $t$. When $\rho=\infty$\,, the
solution is sitting at the north pole ($\theta=0$). 
As $\rho$ decreases, $\theta$ increases.  
At $\rho=\rho_{\rm min}$\,, it comes to the turning point, where $\theta$ takes its
maximal value. It is symmetric with respect to this point.
From the ansatz $\phi=t$, it is rotating in S$^5$\,.

\medskip 

Each point on the curves 
in Figs.\,\ref{dGW_global}-(a), (b) and (c) 
corresponds to a three-dimensional space
parametrized by $(t,\varphi_1,\varphi_2)$.
Each ${\rm S}^2$ parametrized by $(\varphi_1,\varphi_2)$ is centered at
the horizontal axis. Its radius is given by $L \sinh \rho \sin \chi$ and
greater than $LC_2$\,. 
Hence, when $C_2$ is kept finite, 
the radius of the ${\rm S}^2$ 
is much larger than the string length in the large $\lambda$ limit.

\medskip 

Figure \ref{dGW_global}-(c) shows a typical behavior of the solution 
with $C_1 > 1$\,. For a small value of $C_2$\,, the solution looks like 
a dual giant graviton with thin spikes sticking out of the north and 
the south poles. As $C_2$ increases, the radius of the spike becomes larger 
and the shape of dual giant graviton tends to be indistinguishable. 
Also in the case with $C_1 \leq 1$\,, it is hard to find 
the dual giant graviton even for small values of $C_2$ 
as shown in Figs.\,\ref{dGW_global}-(a) and (b).

\medskip

\begin{figure}[tb]
\begin{center}
\begin{minipage}{6cm}
\begin{center}
\includegraphics[width=4cm,angle=270]{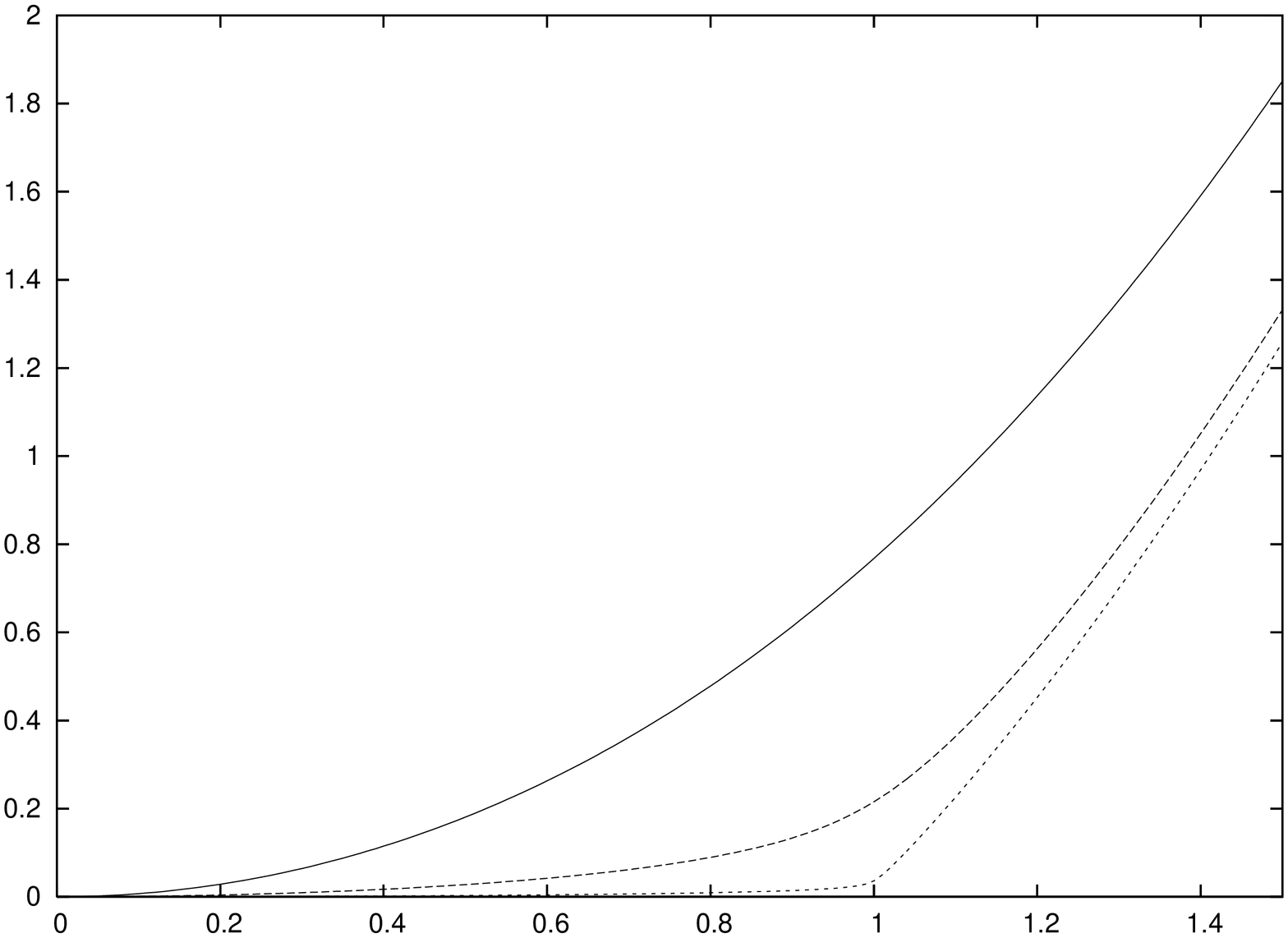}
\put(-170,-58){{\tiny $ J \over N $}}
\put(-85,-120){{\tiny $C_1$}}
\\[2mm]
{\footnotesize (a)}
\end{center}
\end{minipage} \hspace{1cm}
\begin{minipage}{6cm}
\begin{center}
\includegraphics[width=4cm,angle=270]{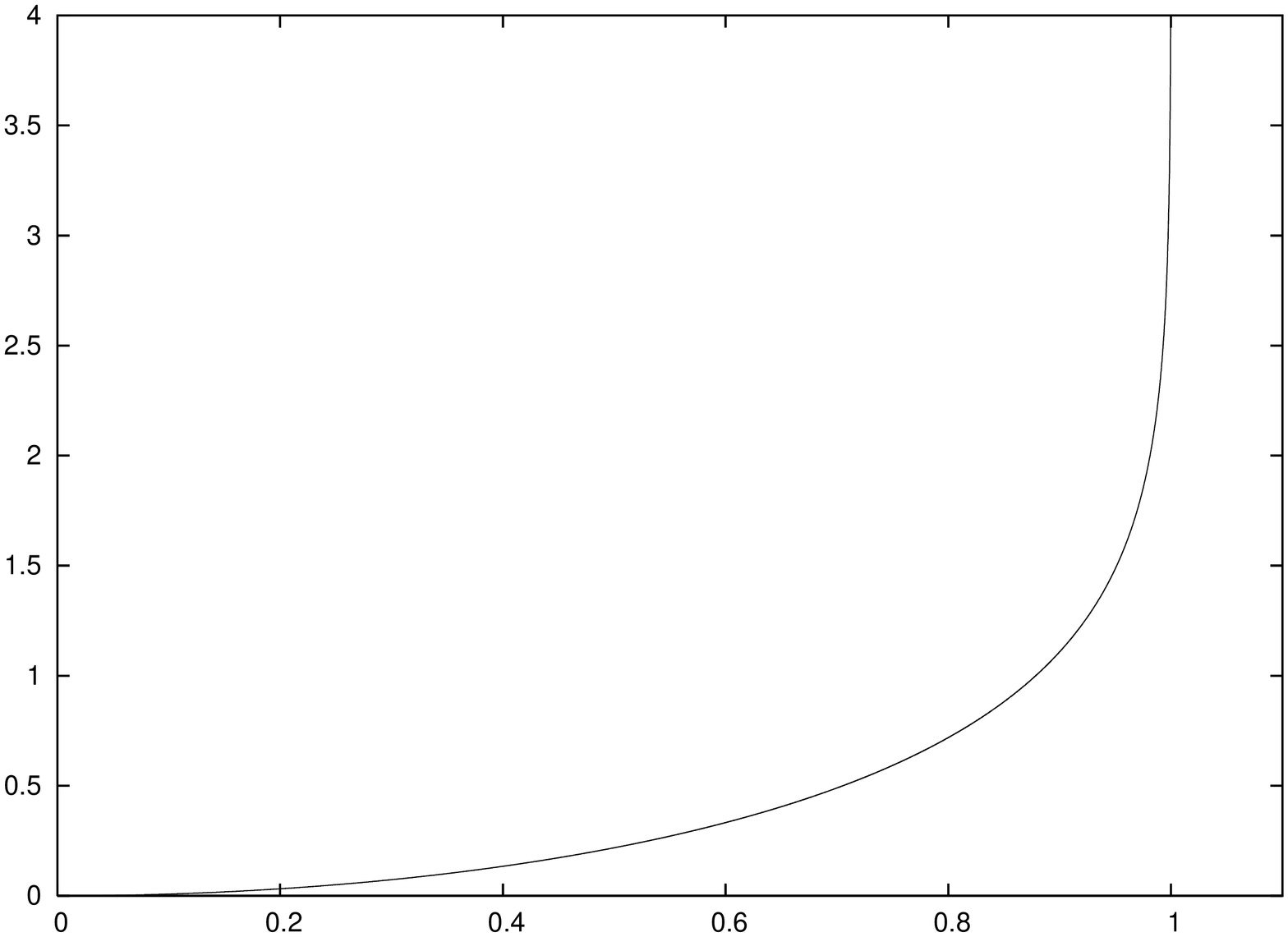}
\put(-178,-58){{\tiny ${ J \over k \sqrt{\lambda} }$}}
\put(-85,-120){{\tiny $C_1$}}
\\[2mm]
{\footnotesize (b)}
\end{center}
\end{minipage}
\end{center}
\caption{\footnotesize The values of $J/N$ and $J/k \sqrt{\lambda}$ 
as functions of $C_1$.}
\label{J_fig}
\end{figure}

We shall examine the behavior of $J$ as a function of $C_1$ 
for a fixed value of $C_2$\,.
Each curve in Fig.\,\ref{J_fig}-(a) shows 
the behavior of $J/N$ with $C_2=1$, $0.1$ and $0.01$ from the top down.
In the limit $C_2 \to 0$\,, the curve asymptotically approaches 
the following line:
\begin{equation}
{J \over N} = 0 \quad ( C_1 \leq 1 )\,, \qquad
{J \over N} = C_1^2 - 1 = \sinh^2 \rho_{\rm min} \quad ( C_1 \geq 1 )\,. 
\end{equation}
The curve $J/N = \sinh^2 \rho_{\rm min}$ corresponds to the 
angular momentum of the dual giant graviton whose 
${\rm S}^3$-radius is given by $\sinh \rho_{\rm min} = \sqrt{C_1^2-1}$ \cite{dGG1,dGG2}.
Figure \ref{J_fig}-(b) depicts the ratio $J/k\sqrt{\lambda}$ 
in the limit $C_2 \to 0$\,. 

\medskip

For finite $k$ the radius of each S$^2$ becomes much smaller than the
string length for the region $\rho > {\rm arccosh}C_1$\,.
Hence the analysis with DBI action may not be reliable because
of possible $\alpha'$-corrections. 
However, a reasonable result has been obtained 
even for a single string case $k=1$ 
and thus the DBI analysis seems to work well even in this case\footnote{
In particular, by setting $k=1$ and $C_1=1$ the D3-brane solution
formally reproduces the string solution in the previous section.}. 
Now we have no obvious reason to believe it 
but guess that the corrections cancel 
each other possibly due to the supersymmetries preserved by the solution.

\subsection{Dual giant Wilson loop around tunneling trajectory}
\label{tunnelingdGW}

Let us now discuss the double Wick rotation 
and the coordinate transformation \eqref{trans_2} 
for the rotating D3-brane solution given by \eqref{D3_chi_theta} 
and \eqref{D3_F}. 

\medskip 

First let us consider the Wick rotation. Here note that 
an imaginary electric flux $F_{t_{\rm E} \rho} 
= - i (L^2 / 2 \pi \alpha') F(\rho)$ should be 
considered in addition to the double Wick rotation 
$t_{\rm E}=it$ and $\phi_{\rm E}=i\phi$\,. 
Then the solution is given by 
\begin{align}
 &
 \sin \chi(\rho) 
 = 
 {C_2 \coth \rho \over \sqrt{\cosh^2 \rho - C_1^2}}\,,\quad
 \sin \theta(\rho) = {C_1 \over \cosh \rho}\,, \quad
 \phi_{\rm E} = t_{\rm E}\,, \label{Esol1}\\
 & F(\rho) = 
 - 
 { \cosh^4 \rho - C_1^2  
   \over 
   \cosh^2 \rho \sqrt{\cosh^2 \rho - C_1^2 -C_2^2 \coth^2 \rho}}\,.
 \label{Esol2}
\end{align}
Next the solution is mapped via \eqref{trans_2} 
and then we have the following D3-brane solution
in the Euclidean Poincar\'e coordinate:
\begin{align}
& 
Z={ \ell \over f }, \quad 
X_1 = 
{ \ell \over f } 
\sinh \rho \sin \chi(\rho) \sin \varphi_1 \cos \varphi_2\,, \quad
X_2 = 
{ \ell \over f } 
\sinh \rho \sin \chi(\rho) \sin \varphi_1 \sin \varphi_2\,, 
\notag 
\\
&
X_3 
= 
{ \ell \over f } 
\sinh \rho 
( 
\alpha \sin \chi(\rho) \cos \varphi_1 
- 
\sqrt{1-\alpha^2} \cos \chi(\rho) 
)\,, \quad
X_4 
= 
{ \ell \over f } \sinh t_{\rm E} \cosh \rho\,, \label{EPSolution}\\
& 
f = 
\cosh t_{\rm E} \cosh \rho 
+ 
\sinh \rho 
( 
\sqrt{1-\alpha^2} \sin \chi(\rho) \cos \varphi_1 + \alpha \cos \chi(\rho) 
)\,. \notag  
\end{align}
Now the function $\chi(\rho)$ is defined by 
the first equation of \eqref{Esol1}.
The shape of the solution is numerically plotted for some values of 
$C_1$ and $C_2$ in Figs.\,5 and 6. 

\medskip 

In order to see the relation to the Wilson loop, 
we shall examine the boundary behavior of the solution \eqref{EPSolution}
by taking the limit $\rho \to \infty$\,. 
Then, for $\alpha \neq 1$\,, the boundary of the solution is
given by the following trajectory on the AdS boundary:
\begin{equation}
Z=X_1=X_2=0\,, \quad
X_3={\mp \ell \sqrt{1-\alpha^2} \over \cosh t_{\rm E} \pm \alpha }\,, \quad
X_4={\ell \sinh t_{\rm E} \over \cosh t_{\rm E} \pm \alpha}\,. 
\label{boundary_C} 
\end{equation} 
The upper (the lower) sign implies the region with
$0 \leq \chi \leq \pi/2$ ($\pi/2 \leq \chi  \leq \pi$). This is a circle with 
the radius $\ell / \sqrt{1-\alpha^2}$ on the $(X_3, X_4)$-plane. 
Its center is located at $(X_3,X_4)=({\alpha
\ell / \sqrt{1 - \alpha^2}},0)$\,. 

\medskip 

For $\alpha=1$\,, the circle \eqref{boundary_C}, except for $t_{\rm E}=0$\,,  
becomes an infinite line, $X_3=0$\,.
That is, the D3-brane is attached to a straight line 
on the AdS boundary and extended infinitely into the bulk AdS space. 
This infinitely extended part of the D3-brane 
can be found 
by carefully considering $t_{\rm E}=0$\,. 
For example, let us take the large $\rho$ limit of the solution \eqref{EPSolution}
after setting $t_{\rm E}=0$\,. Then we reach the AdS
boundary ($Z=0$) on the patch with $0 \leq \chi \leq \pi/2$\, 
but we go to the region at the vicinity of $Z \sim \infty$ 
on the other patch $\pi/2 \leq \chi \leq \pi$\,. 
It is easy to check that, for $\alpha=1$\,, 
$\sqrt{X_1^2+X_2^2+X_3^2}/Z \to C_2$ in the large $\rho$ limit. 
This means that the solution asymptotically
satisfies the linear ansatz used in \cite{Drukker:2005kx}.

\medskip

\begin{figure}[tb]
\begin{center}
\begin{minipage}{4cm}
\includegraphics[width=4cm]{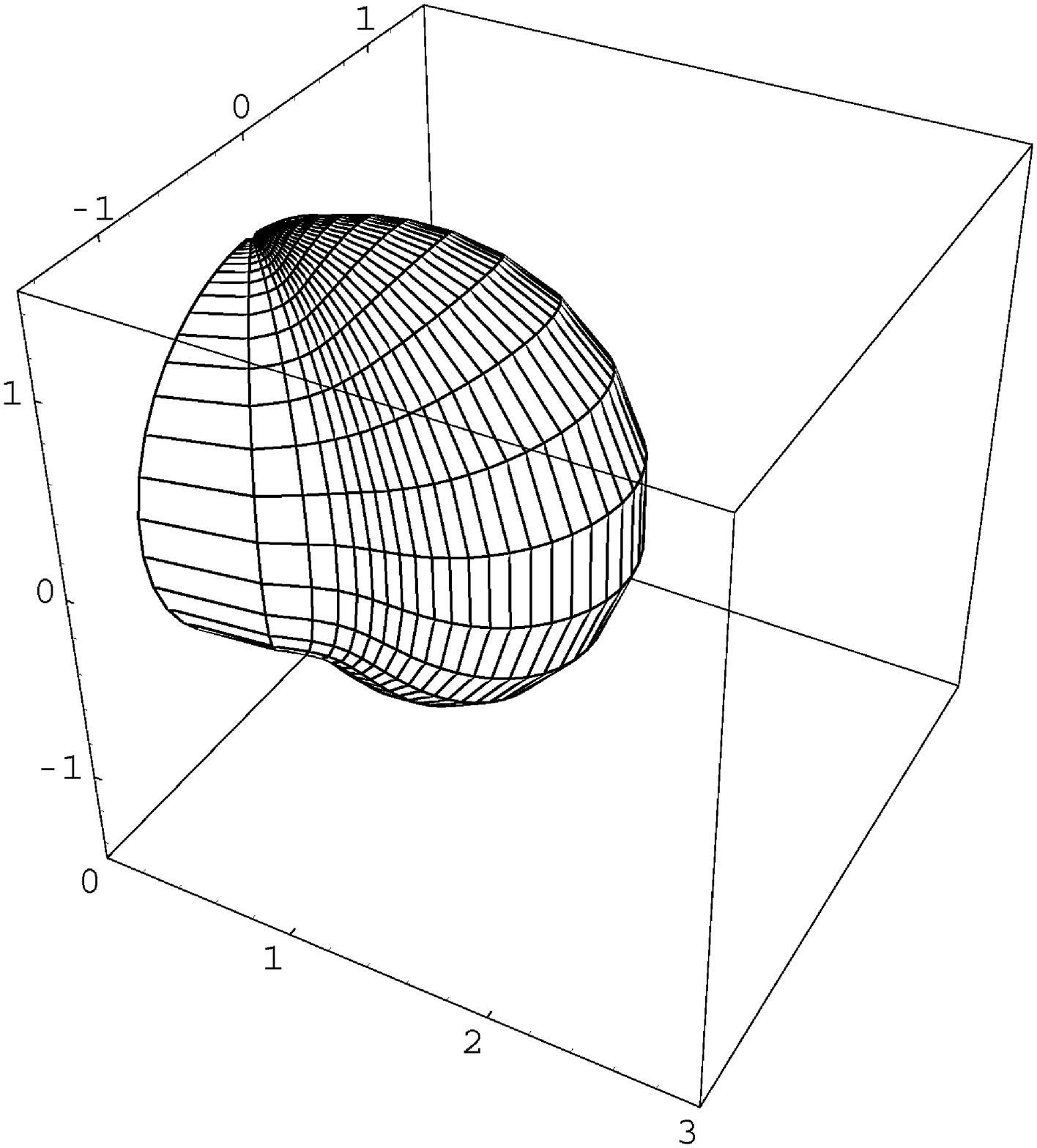}
\put(-85,10){\scriptsize $Z/\ell$}
\put(-110,120){\scriptsize $X_3/\ell$}
\put(-135,60){\scriptsize $X_4/\ell$}
\\ \hspace*{-0.7cm} 
{\footnotesize (a)\quad $(C_1,C_2,\alpha)=(1.2,0.1,0)$}
\end{minipage} \hspace{2.5cm}
\begin{minipage}{4.1cm}
\includegraphics[width=4cm]{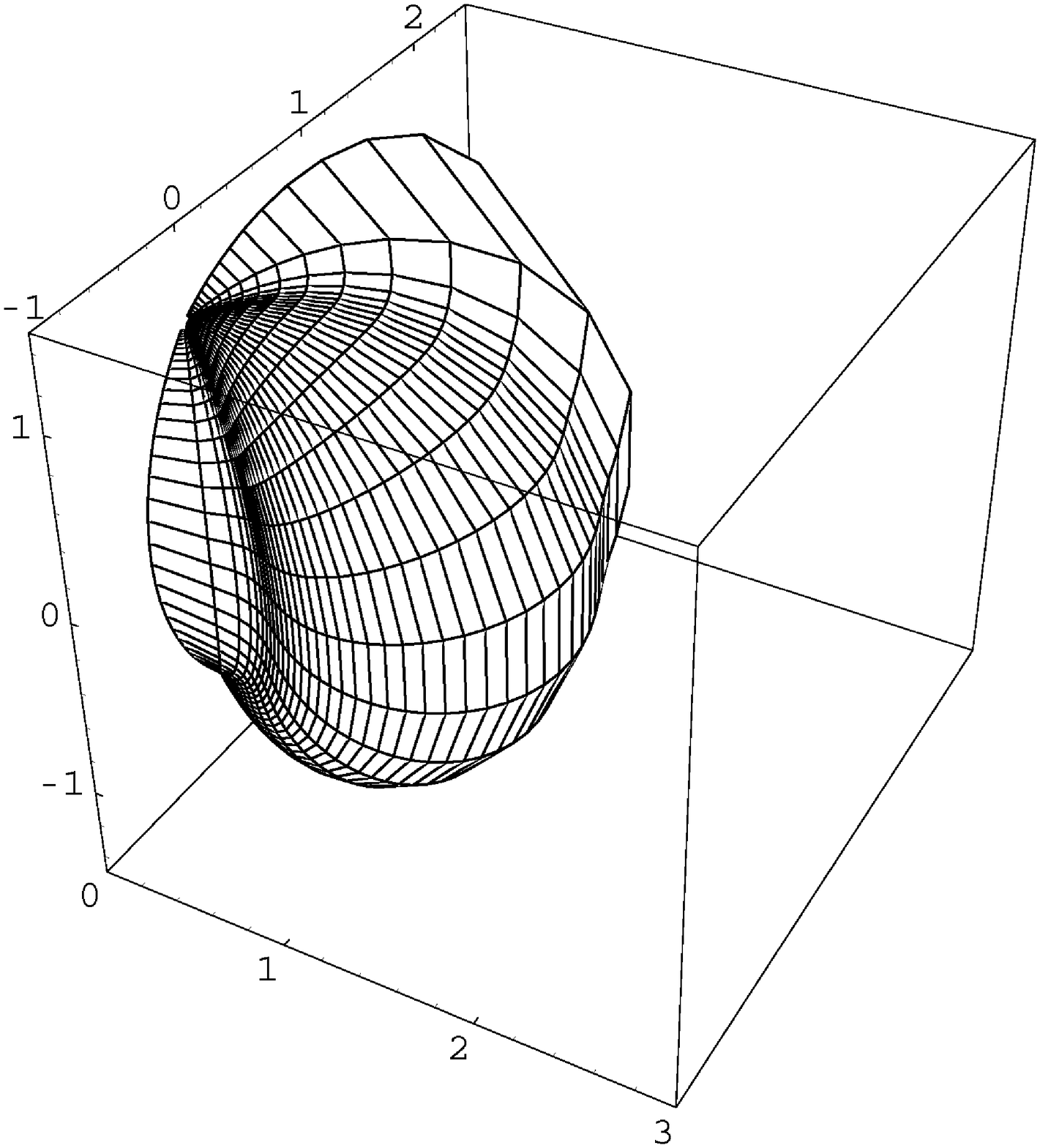}
\put(-86,8){\scriptsize $Z/\ell$}
\put(-110,116){\scriptsize $X_3/\ell$}
\put(-133,58){\scriptsize $X_4/\ell$}
\\ \hspace*{-0.9cm} 
{\footnotesize (b)\quad $(C_1,C_2,\alpha)=(1.2,0.1,0.7)$}
\end{minipage} \\[8mm]
\begin{minipage}{4cm}
\includegraphics[width=4cm]{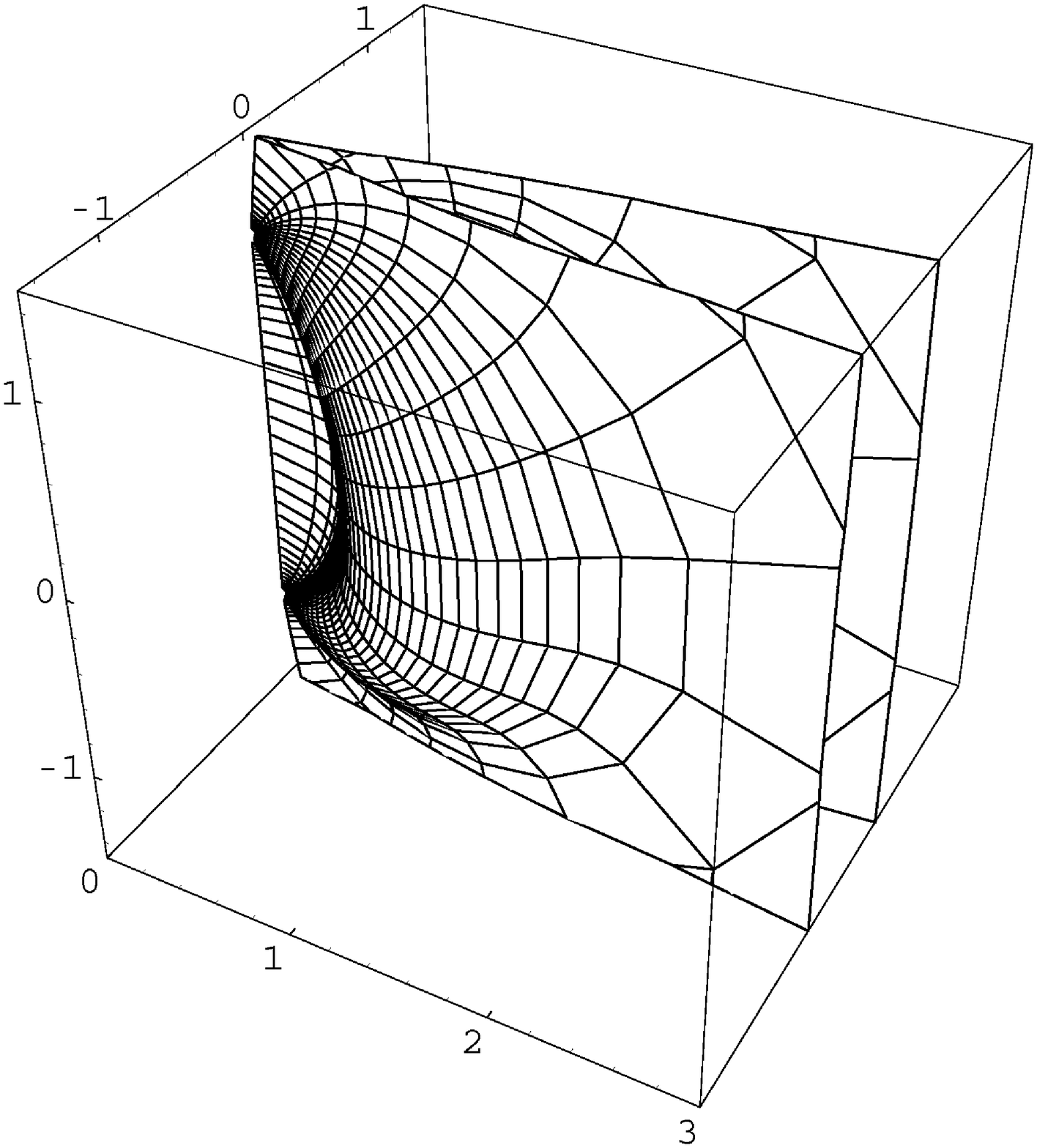}
\put(-85,10){\scriptsize $Z/\ell$}
\put(-110,120){\scriptsize $X_3/\ell$}
\put(-135,60){\scriptsize $X_4/\ell$}
\\ \hspace*{-0.7cm} 
{\footnotesize (c)\quad $(C_1,C_2,\alpha)=(1.2,0.1,1)$} 
\end{minipage} \hspace{2.5cm}
\begin{minipage}{4cm}
\includegraphics[width=4cm]{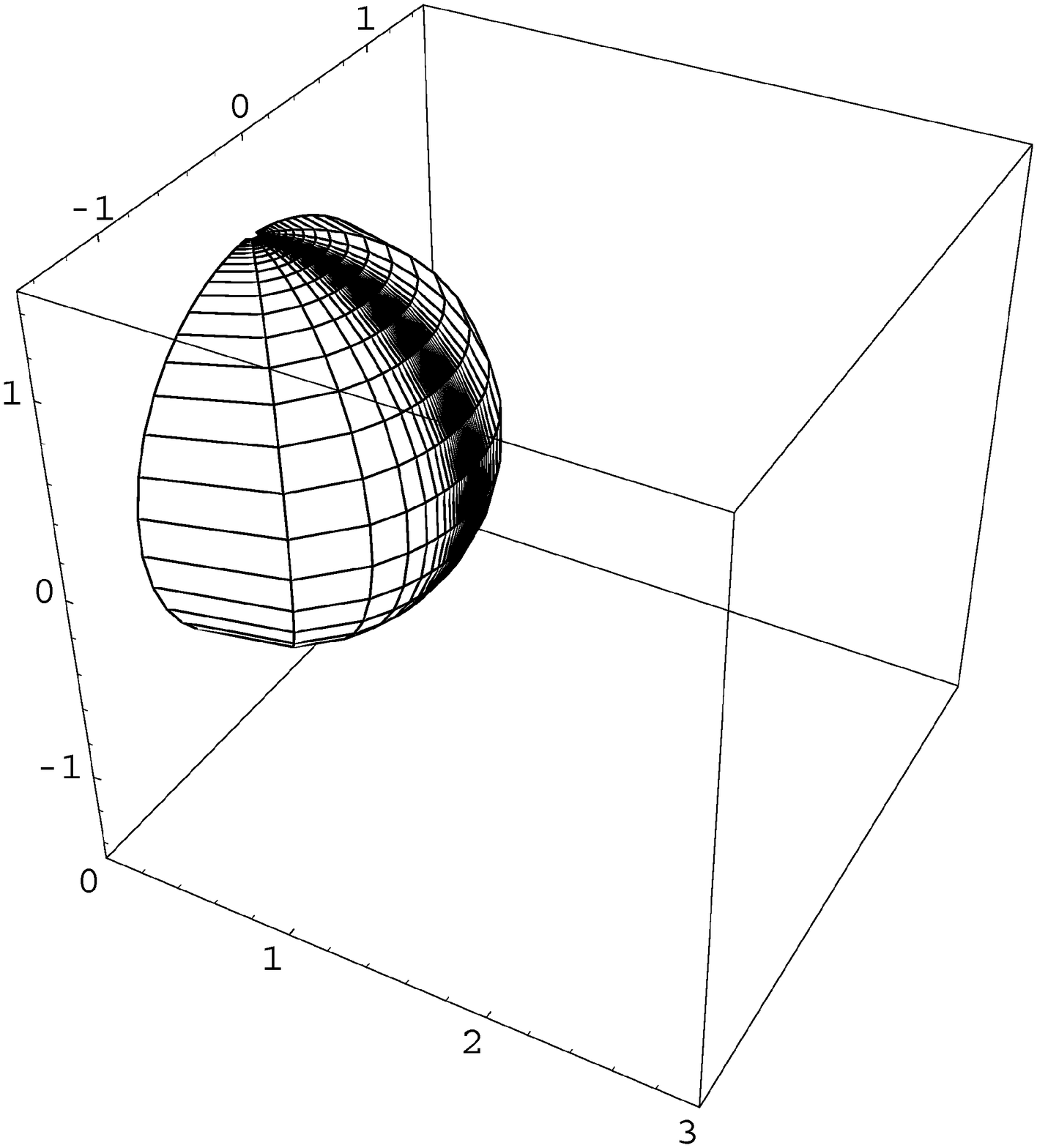}
\put(-85,10){\scriptsize $Z/\ell$}
\put(-110,120){\scriptsize $X_3/\ell$}
\put(-135,60){\scriptsize $X_4/\ell$}
\\ \hspace*{-0.7cm} 
{\footnotesize (d)\quad $(C_1,C_2,\alpha)=(0,0.1,0)$} 
\end{minipage}
\end{center}
\caption{\footnotesize 
Euclidean D3-brane solution with various values of $C_1$, $C_2$ 
and $\alpha$\,. Figure (d) corresponds to the non-rotating Drukker-Fiol 
solution \cite{Drukker:2005kx}.
}  \label{dGW3D}
\end{figure}

\begin{figure}[tb]
\begin{center}
\begin{minipage}{4cm}
\includegraphics[width=4cm]{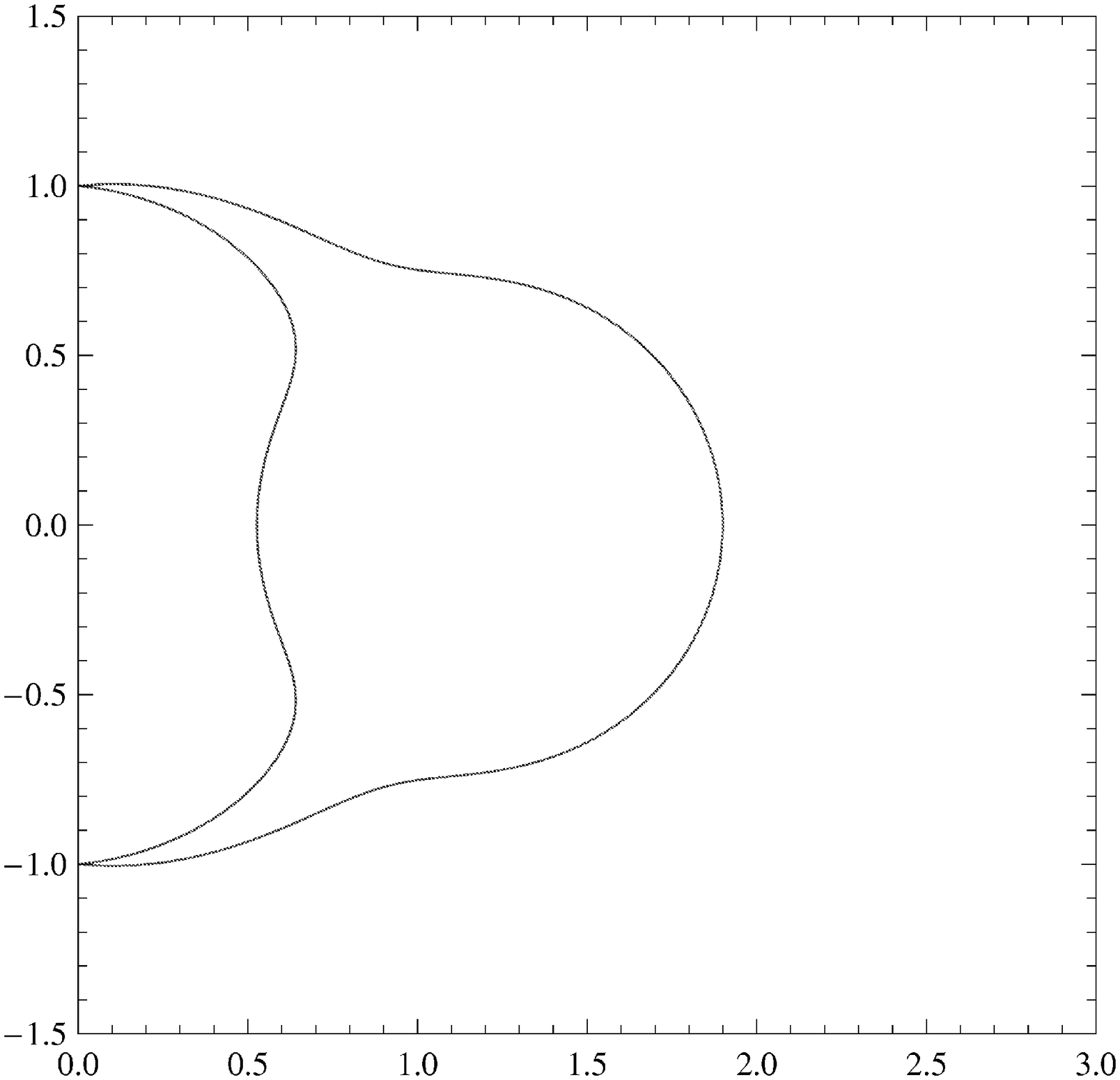}
\put(-67,-10){\scriptsize $Z/\ell$}
\put(-135,55){\scriptsize $X_3/\ell$}
\vspace*{0.5cm}\\ \hspace*{-0.7cm} 
{\footnotesize (a)\quad $(C_1,C_2,\alpha)=(1.2,0.1,0)$} 
\end{minipage} \hspace{2.0cm}
\begin{minipage}{4.1cm}
\includegraphics[width=4cm]{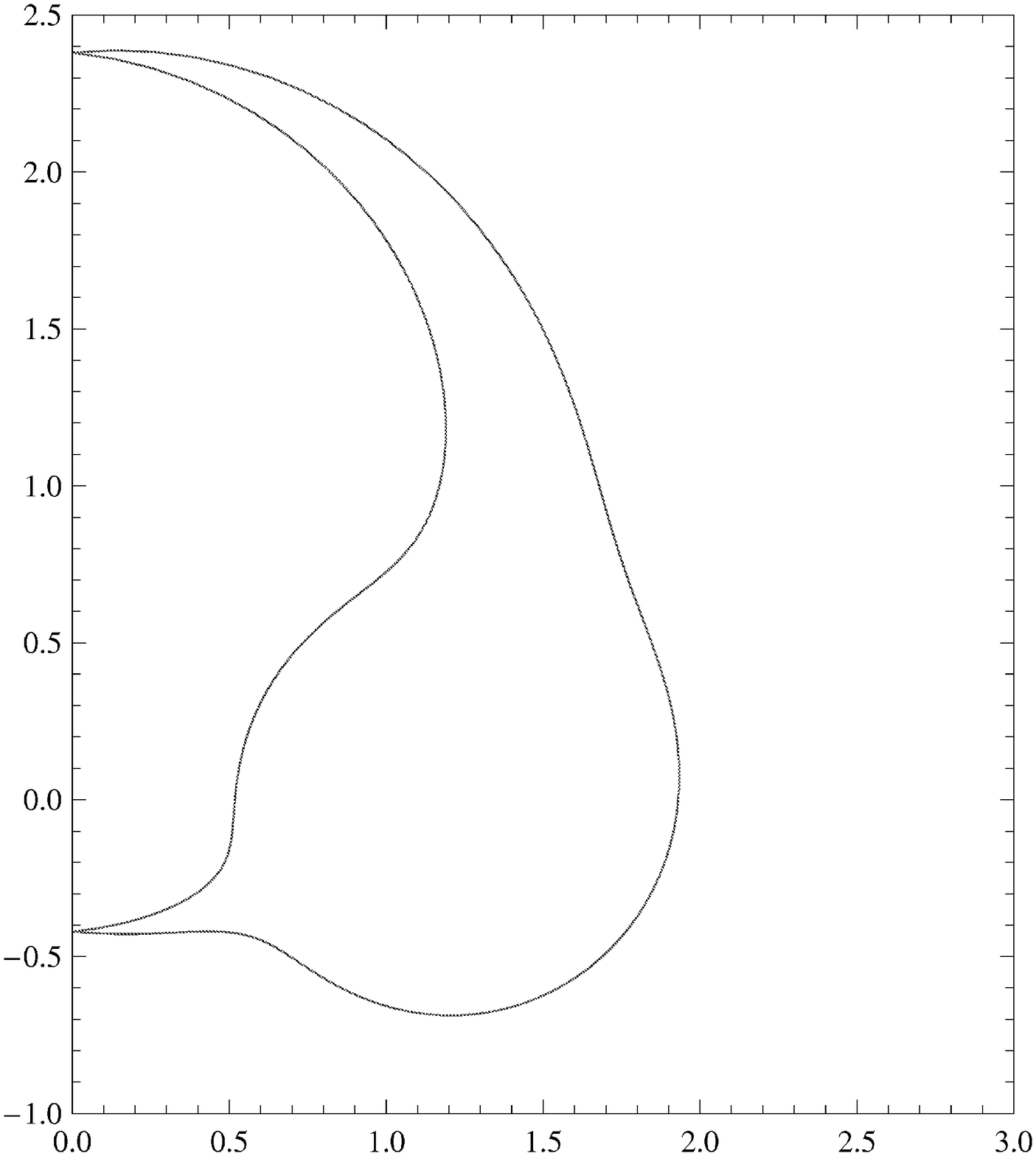}
\put(-67,-10){\scriptsize $Z/\ell$}
\put(-135,65){\scriptsize $X_3/\ell$}
\\ \hspace*{-0.9cm} 
{\footnotesize (b)\quad $(C_1,C_2,\alpha)=(1.2,0.1,0.7)$} 
\end{minipage} \\[4mm]
\begin{minipage}{4cm}
\includegraphics[width=4cm]{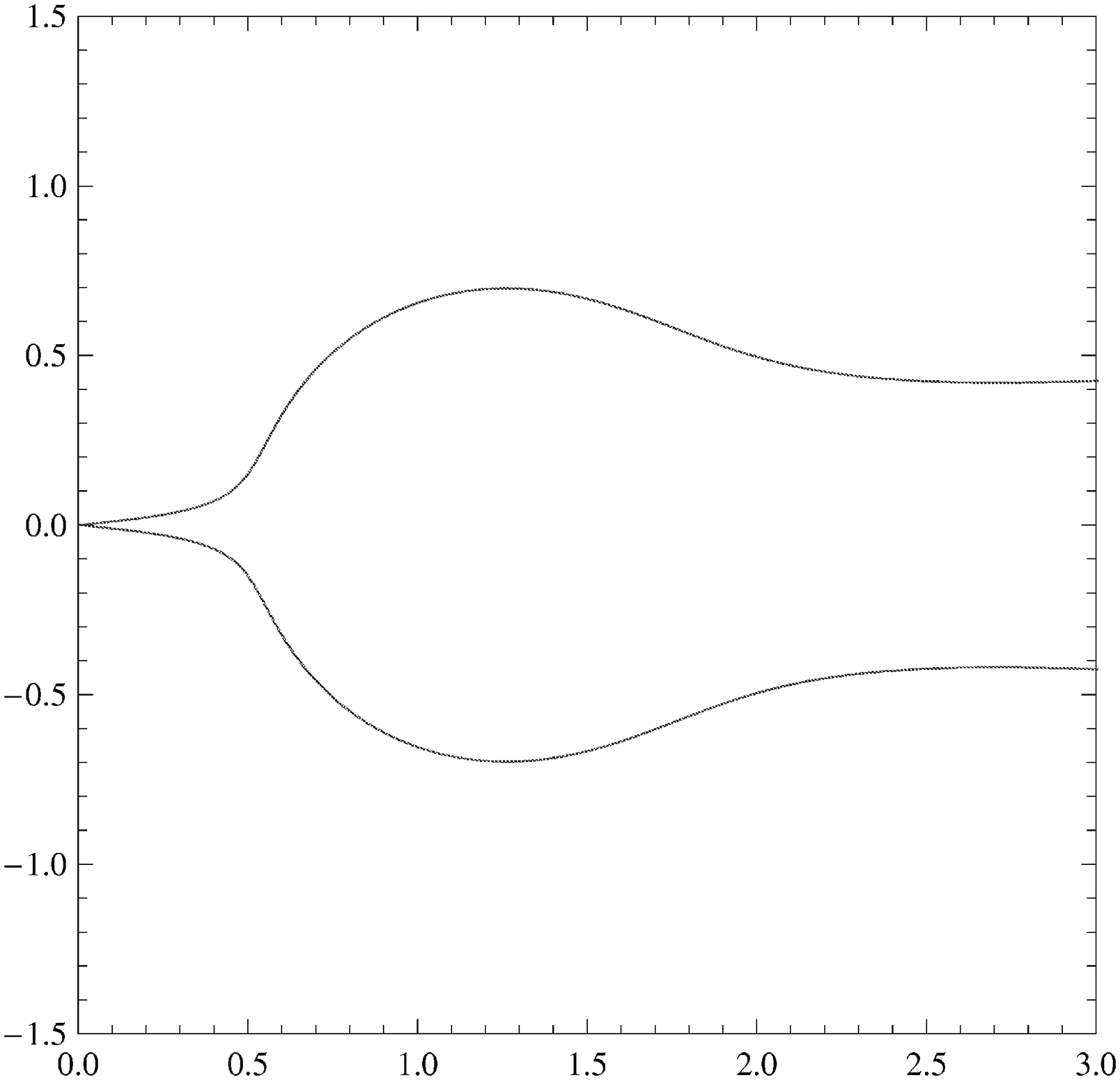}
\put(-67,-10){\scriptsize $Z/\ell$}
\put(-135,55){\scriptsize $X_3/\ell$}
\vspace*{0.12cm} \\ \hspace*{-0.7cm} 
{\footnotesize (c)\quad $(C_1,C_2,\alpha)=(1.2,0.1,1)$} 
\end{minipage} \hspace{2.0cm}
\begin{minipage}{4cm}
\includegraphics[width=4cm]{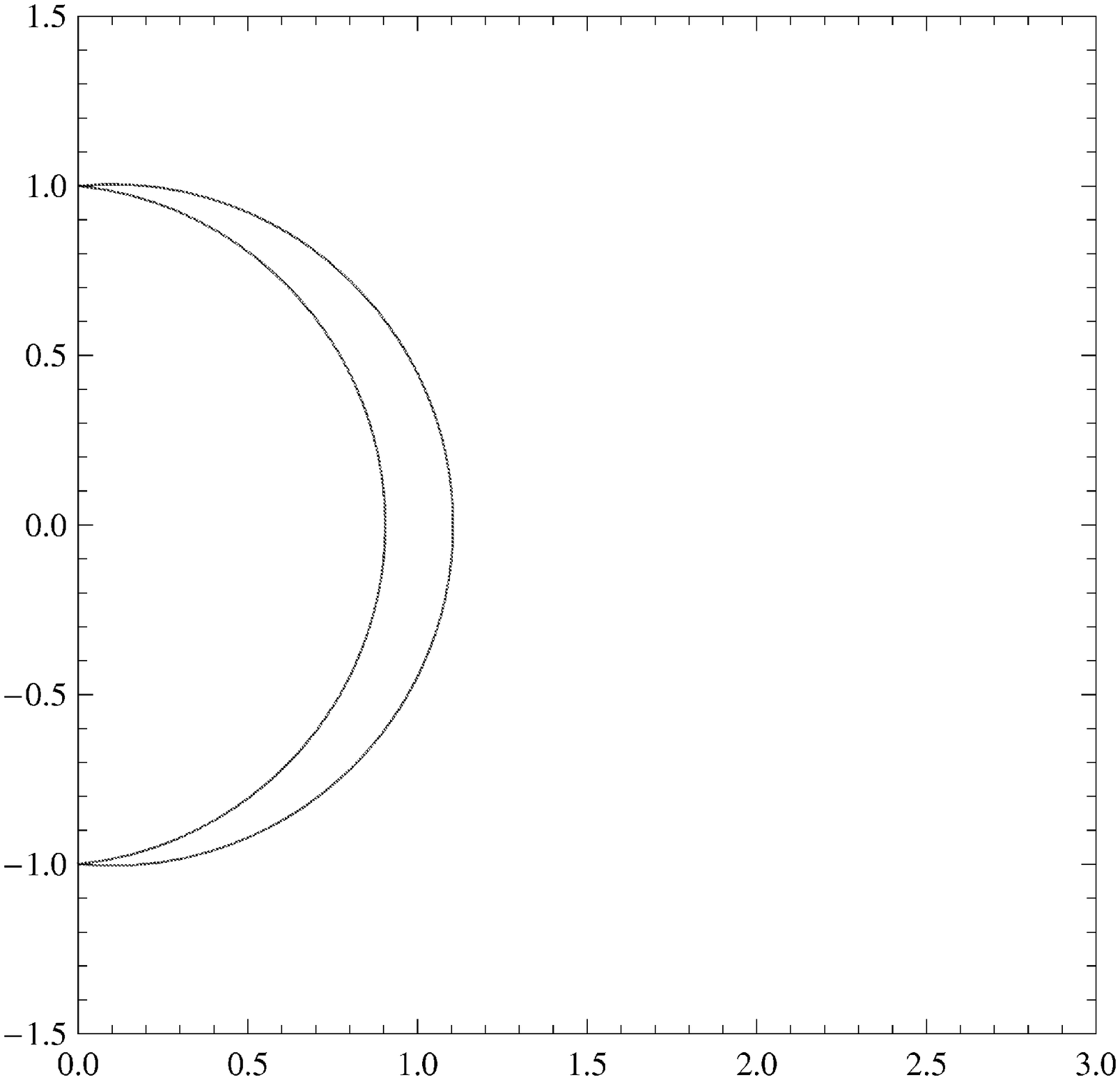}
\put(-67,-10){\scriptsize $Z/\ell$}
\put(-135,55){\scriptsize $X_3/\ell$}
\vspace*{0.12cm}\\ \hspace*{-0.7cm} 
{\footnotesize (d)\quad $(C_1,C_2,\alpha)=(0,0.1,0)$} 
\end{minipage} \\[4mm]
\end{center}
\caption{\footnotesize The time slices of Fig.\,\ref{dGW3D} at $t=0$\,.}
\label{dGW2D}
\end{figure}

Figures \ref{dGW3D} and \ref{dGW2D} are some numerical plots 
of the D3-brane solution 
with indicated values of $C_1$, $C_2$ and $\alpha$\,.
Figure \ref{dGW3D} depicts two-dimensional surfaces 
specified by $\sin \varphi_1=0$,
and Fig.\,\ref{dGW2D} expresses their cross sections at $X_4=0$.  
In particular, Figs.\,\ref{dGW3D}-(d) and \ref{dGW2D}-(d) 
correspond to the non-rotating Drukker-Fiol solution \cite{Drukker:2005kx}.

\medskip 

From Figs.\,\ref{dGG3D}, \ref{dGW3D} and \ref{dGW2D},
it is manifestly observed again that the solution with $(C_1,C_2)=(1.2,0.1)$ 
is composed of a dual giant graviton propagating along 
the tunneling trajectory and a spike D3-brane solution.
For the values of $C_1$ and $C_2$\,, 
the presence of the dual giant graviton is obvious.
As $C_2$ increases or $C_1$ decreases, the spike tends 
to be wider compared to the radius of the dual giant graviton 
and absorbs it.  

\medskip 

Thus the solution \eqref{EPSolution} is attached to
a circle or a straight line on the Poincar\'e AdS boundary and it is
carrying an angular momentum $J$ from a point on the boundary to another.

\medskip

Finally we shall give a comment on the operator 
corresponding to the Euclidean D3-brane solution.
As we have already explained, 
a natural candidate should be a circular or a straight-line Wilson loop 
in $k$-th symmetric representation with local operator insertions.
However, for the solutions with $J$ of order $N$ or larger, 
it may be necessary to take account of non-planar contributions 
for the local operators. That is, we may have to replace $Z^J$ 
with the dual giant graviton operator like \eqref{dGG_dGG}.
In fact, in the dual gravity side, 
the rotating D3-brane is composed of a dual giant Wilson 
loop and a dual giant graviton. 
We will further discuss the corresponding gauge-theory operator again 
in section \ref{Wilson_Loop}. 

\section{Evaluation of D3-brane action}
\label{Evaluation}

Let us evaluate the classical action of the Euclidean D3-brane solution. 
For simplicity we omit the subscript ``E'' hereafter. 
All $t$ and $\phi$ in the following should be understood 
as $t_{\rm E}$ and $\phi_{\rm E}$\,. 

\medskip

In addition to the DBI action $S_{\rm DBI}$ and the WZ term $S_{\rm WZ}$, 
we have to add appropriate boundary terms to adjust the boundary
conditions properly.
First of all, it is necessary to consider the usual boundary term
for the Legendre transformation of the radial coordinate $u=1/Z$ of the
${\rm AdS}_5$ \cite{Drukker:1999zq}.
Then we have to introduce additional boundary terms for 
other Legendre transformations because  
the solution carries the conserved charges: 
the string charge $k$ and the angular momentum $J$\,.  

\medskip 

After all, the following summation should be considered as the total action:
\begin{equation}
S_{\rm total} = S_{\rm DBI} + S_{\rm WZ} + S_{\phi} + S_A + S_u\,. 
\label{Routhian} 
\end{equation}
The last three terms are the boundary terms for the Legendre
transformations with respect to the angle variable $\phi$\,, 
the gauge potential $A$ and the radial coordinate $u=1/Z$\,. 

\medskip 

For the solution we consider, all the terms in \eqref{Routhian} contain divergences. Hence it is necessary to introduce cutoffs 
$t_{\rm min}$\,, $t_{\rm max}$ and
$\rho_{\rm max}$, and restrict the range of the integration as
\begin{equation}
t_{\rm min} < t < t_{\rm max}\,, \quad
\rho_{\rm min} < \rho < \rho_{\rm max}\,, \quad
0 \leq \varphi_1 \leq \pi\,, \quad
0 \leq \varphi_2 \leq 2 \pi\,. \nonumber 
\end{equation}
Remember that $\rho_{\rm min}$ is defined by 
$\sin \chi ( \rho_{\rm min} ) = 1$\,, or $\rho_{\rm min}=0$ 
for the solution with $\chi(\rho)=0$\,.

\medskip 

We use two notations of the world-volume coordinate hereafter.  
The one is the notation we used so far, and  
$\rho$ is regarded as a world-volume coordinate. 
Then we have to consider the two regions $0\leq \chi \leq
\pi/2$ and $\pi/2 \leq \chi \leq \pi$\,. 
The other is to use $\chi$ as a
world-volume coordinate, instead of $\rho$\,. Then the whole solution
can be covered with a single patch. Hereafter we shall occasionally use
this single-patch notation, where the range of the parameter $\chi$ is
restricted as $\chi_{\rm min} < \chi < \chi_{\rm max}$ with $\chi_{\rm
min}=\chi(\rho_{\rm max})$ and $\chi_{\rm max} = \pi - \chi(\rho_{\rm
max})$\,.\footnote{
Note that, for the solution with $\chi(\rho)=0$\,, 
we need to take $\rho$ as a world volume coordinate.
}  

\medskip

From now on we evaluate the $\alpha$- and $\ell$-dependence
of each term in \eqref{Routhian}. To make our discussion clear, 
we shall summarize below the relevant steps of the calculation 
and the results only. We refer the readers, who are interested 
in the detailed calculations, to Appendices. 

\subsection{Evaluation of $ \boldsymbol{ S_{\rm DBI} + S_{\rm WZ} }$ }

The aim here is to evaluate the contributions coming from the
DBI action and the WZ term:
\begin{align}
S_{\rm DBI} + S_{\rm WZ} 
& = 
\int\!\! dt d\rho d \varphi_1 d \varphi_2\, {\cal L} \notag \\
& =
T_{\rm D3} \int\!\! dt d\rho d \varphi_1 d \varphi_2\, 
\sqrt{\det(G_{ab} + 2 \pi \alpha' F_{ab})}
- T_{\rm D3} \int {\cal P}_\alpha [{\cal C}_4]\,.
\end{align}
Recall that there are contributions from two patches, though it is 
not written down explicitly. Our notation will be explained shortly.

\medskip

Remember that the solution \eqref{EPSolution}
should be derived from the action defined on the double Wick rotated geometry:
\begin{equation}
 { ds^2 \over L^2 } 
 = 
 {dZ^2 + (dX_1)^2 + (dX_2)^2 + (dX_3)^2 + (dX_4)^2 \over Z^2} 
 + d \theta^2 - \sin^2 \theta d \phi^2\,,
\end{equation}
and the imaginary ansatz for the electric flux:
\begin{equation}
 F_{t \rho} = - i { L^2 \over 2 \pi \alpha'} F\,. 
\end{equation}

\medskip 

The explicit form of $S_{\rm DBI}$ is 
\begin{equation}
S_{\rm DBI}
=
2 \times {2 N \over \pi}
\int_{t_{\rm min}}^{t_{\rm max}} dt
\int_{\rho_{\rm min}}^{\rho_{\rm max}} d\rho 
{ C_2^3 ( \cosh^4 \rho - C_1^2 ) \cosh^2 \rho 
\over
( \cosh^2 \rho - C_1^2 )^2 
\sqrt{ \cosh^2 \rho - C_1^2 - C_2^2 \coth^2 \rho }
}\,. \label{DBI}
\end{equation}
The overall factor $2$ implies that there are two patches. 
Possible dependence on $\alpha$ and $\ell$ arises only through 
the definition of the cutoffs.

\medskip 

As for the WZ term, there is an ambiguity related to the gauge
transformation: Under the gauge transformation the RR potential may
change by an exact form, and it may affect the value of the WZ term
since boundaries of the D3-brane should be taken into account. 

\medskip 

In order to fix this ambiguity, we just follow the proposal of
\cite{Drukker:2006zk} and use the following RR potential,
\begin{equation}
{\cal C}_4 = { L^4 \over Z^4 } 
dX_4 \wedge dX_1 \wedge dX_2 \wedge dX_3\,.
\label{RR}
\end{equation}
The pull back of the RR potential on the D3-brane solution is
represented by ${\cal P}_\alpha[{\cal C}_4]$ and defined on the space
spanned by $(t,\chi,\varphi_1,\varphi_2)$\,.\footnote{ We use the
single-patch notation to discuss the WZ term.}   
As it can be seen from \eqref{EPSolution}\,, 
the pull back depends
on $\alpha$ but it is independent of $\ell$\,.
The solutions with
different values of $\alpha$ are related through coordinate
transformation. 
This is the case for the pull backs with different 
$\alpha$\,,
and the WZ term  
\begin{equation}
S_{\rm WZ} 
= - T_{\rm D3} \int {\cal P}_\alpha [{\cal C}_4]\,
\label{S_WZ^a}
\end{equation}
depends on $\alpha$ through the boundary term. 

\medskip 

In order to evaluate the WZ term, it is convenient to introduce a three
form $\Lambda_3^\alpha$ as follows:
\begin{equation}
 {\cal P}_\alpha [{\cal C}_4] - {\cal P}[\widetilde C_4] = d \Lambda_3^\alpha\,.
 \label{P-P}
\end{equation}
Here ${\cal P}[\widetilde C_4]$ is written in terms of 
$(t,\chi,\varphi_1,\varphi_2)$ as 
\begin{equation}
 {\cal P} [\widetilde C_4] = L^4 \sinh^4 \rho (\chi)\, \sin^2 \chi\, \sin \varphi_1\, 
 dt \wedge d \chi \wedge d \varphi_1 \wedge d \varphi_2\,, \label{4.7}
\end{equation}
where $\rho(\chi)$ is the inverse function of $\chi=\chi(\rho)$ (and
also $\chi=\pi-\chi(\rho)$) and the explicit form of $\Lambda_3^\alpha$ is
given in Appendix \ref{C4_L3}. 
Note that the DBI action \eqref{DBI}
can be written as\footnote{
This just means the fact that $S_{\rm DBI} + S_{\rm WZ}=0$
in the original global coordinate.
} 
\begin{equation}
 S_{\rm DBI} = T_{\rm D3} \int {\cal P}[\widetilde C_4]\,.  
\label{S_DBI+S_WZ=0}
\end{equation}
By using \eqref{S_WZ^a}, \eqref{P-P} and \eqref{S_DBI+S_WZ=0},  
we obtain the following expression, 
\begin{equation}
S_{\rm DBI} + S_{\rm WZ}
= - T_{\rm D3} \int_{\rm b} \Lambda_3^\alpha\,.
\label{DBI+WZ}
\end{equation}
Here the subscript ``b'' implies that the integral is over the
boundary of the space parametrized by $(t,\chi,\varphi_1,\varphi_2)$\,. 

\medskip 

The non-vanishing components
of $\Lambda_3^\alpha$ are as follows:
\begin{equation}
\Lambda_3^\alpha
=
(\Lambda_3^\alpha)_{t \chi \varphi_2} 
dt \wedge d \chi \wedge d \varphi_2
+ 
(\Lambda_3^\alpha)_{t \varphi_1 \varphi_2} 
dt \wedge d \varphi_1 \wedge d \varphi_2
+
(\Lambda_3^\alpha)_{\chi \varphi_1 \varphi_2} 
d \chi \wedge d \varphi_1 \wedge d \varphi_2\,. \nonumber 
\end{equation}
The explicit forms are given by \eqref{LtL} with
\eqref{Lambda3''} and \eqref{D3_1}--\eqref{D3_5}.
Then the right-hand side 
of \eqref{DBI+WZ} can be rewritten as 
\begin{align}
\int_{\rm b} \Lambda_3^\alpha 
& =
\int_{t_{\rm min}}^{t_{\rm max}}\!\!\! dt 
\int_{\chi_{\rm min}}^{\chi_{\rm max}}\!\!\! d\chi
\int_0^{2 \pi} \!\!\! d \varphi_2 
\Big[ ( \Lambda_3^\alpha )_{t \chi \varphi_2} 
\Big]^{\varphi_1=\pi}_{\varphi_1=0}
-
\int_{t_{\rm min}}^{t_{\rm max}}\!\!\! dt 
\int_0^\pi \!\!\! d \varphi_1 
\int_0^{2 \pi} \!\!\!d \varphi_2
\Big[ ( \Lambda_3^\alpha )_{t \varphi_1 \varphi_2} 
\Big]^{\chi_{\rm max}}_{\chi_{\rm min}} \notag \\
& \qquad + 
\int_{\chi_{\rm min}}^{\chi_{\rm max}}\!\!\! d \chi
\int_0^{\pi} \!\!\! d \varphi_1
\int_0^{2 \pi} \!\!\! d \varphi_2
\Big[ (  \Lambda_3^\alpha )_{\chi \varphi_1 \varphi_2}
\Big]^{t_{\rm max}}_{t_{\rm min}}\,.
\label{int_P[Lambda3]}
\end{align}
All the integrands in \eqref{int_P[Lambda3]} obviously are independent of 
$\ell$\,. The $\ell$-dependence might arise through the cutoffs $\rho_{\rm
max}$ and $t_{\rm min,max}$\,, but the integrals actually converge 
and do not depend on $\ell$\,. 
The $\alpha$-dependence of the integrals is
discussed in Appendices \ref{C4_L3} and \ref{intL3}. 
Here we just summarize the results:
\begin{enumerate}
\item 1st-term: $(\Lambda_3^\alpha)_{t \chi \varphi_2}$-term \\
This term vanishes thanks to the following relation 
(see \eqref{Lambda3''} and \eqref{D3_1}):
\begin{equation}
 ( \Lambda_3^\alpha )_{t \chi \varphi_2} \Big|_{\varphi_1 = 0}
 =
 ( \Lambda_3^\alpha )_{t \chi \varphi_2} \Big|_{\varphi_1 = \pi}
 =0\,. \nonumber 
\end{equation}

\item 2nd-term: $(\Lambda_3^\alpha)_{t \varphi_1 \varphi_2}$-term \\
A detailed calculation is summarized in Appendix \ref{app_intL3_tvv}. 
Here we rely on a numerical calculation in a step of the 
integrals. The result is as follows: 
\begin{align}
&
\int_{t_{\rm min}}^{t_{\rm max}} dt
\int_0^\pi d \varphi_1 
\int_0^{2 \pi} d \varphi_2
\Big[
(\Lambda_3^\alpha)_{t \varphi_1 \varphi_2}
\Big]_{\chi_{\rm min}}^{\chi_{\rm max}} \notag \\
&\hspace{2cm}
\xrightarrow{|t_{\rm min,max}|\,, \,\rho_{\rm max} \to \infty} \qquad
\begin{cases}
4 \pi^2 L ^4 ({\rm arcsinh C_2} - C_2 \sqrt{1+C_2^2} )
&
(\alpha=1) \\
0
&
(\alpha \neq 1)
\end{cases}
\,. \notag 
\end{align}

\item 3rd-term: $(\Lambda_3^\alpha)_{\chi \varphi_1 \varphi_2}$-term \\
As explained in Appendix \ref{app_intL3_cvv}, this term is independent 
of $\alpha$ in the large $|t_{\rm min,max}|$ limit. The integral is performed 
over the boundary at $t=t_{\rm min,max}$ and it is localized
near the ``points'' where the local operators are inserted. 
Therefore the result should not depend on $\alpha$\,, because 
$\alpha$ describes the global structure of the solution. 

\end{enumerate}

In summary, $S_{\rm DBI} + S_{\rm WZ}$ is given by  
\begin{equation}
S_{\rm DBI} + S_{\rm WZ}
=
\textrm{const.}
+ 
\begin{cases}
2 N \Big(  {\rm arcsinh} C_2  - C_2 \sqrt{1 + C_2^2} \Big) 
&
( \alpha = 1) \\
0 & ( \alpha \neq 1)
\end{cases}
\,. \label{res_SDBI_SWZ}
\end{equation}
Here ``const.'' is a finite constant independent of $\alpha$ and
$\ell$\,, but it may depend on $C_1$ and $C_2$\,.

\subsection{Evaluation of $\boldsymbol {S_\phi + S_A + S_u}$ }

Next we discuss the boundary terms $S_\phi$, $S_A$ and $S_u$\,.

\medskip 

The boundary terms $S_\phi$, $S_A$ and $S_u$ are defined, respectively, as
\begin{align}
 S_\phi 
 &= - \int_{\rho_{\rm min}}^{\rho_{\rm max}} \!\! 
 d\rho d\varphi_1 d \varphi_2\,
\bigg[ 
{ \partial {\cal L} \over \partial \dot \phi } \phi 
\bigg]_{t_{\rm min}}^{t_{\rm max}}\,, 
\label{S_phi} \\
 S_A
 &= - \int_{\rho_{\rm min}}^{\rho_{\rm max}}\!\! d \rho d\varphi_1 d \varphi_2\,
\bigg[
{ \partial {\cal L} \over \partial F_{t \rho} } A_\rho 
\bigg]_{t_{\rm min}}^{t_{\rm max}}\,, 
\label{S_A} \\
 S_u & = -  \int_{\rho_{\rm min}}^{\rho_{\rm max}}\!\! 
 d\rho d\varphi_1 d \varphi_2\,
\bigg[
{\partial {\cal L} \over \partial \dot u} u 
\bigg]_{t_{\rm min}}^{t_{\rm max}}
- \int_{t_{\rm min}}^{t_{\rm max}}\!\! dt d\varphi_1 d \varphi_2\,
{\partial {\cal L} \over \partial u'} u \bigg|_{\rho_{\rm max}}\,. 
\label{S_u}
\end{align}
There are implicitly contributions from the two patches. 

\medskip 

As for \eqref{S_phi} and \eqref{S_A}, the dependence on $\alpha$ and
$\ell$ may come only from the cutoffs.  By
performing $\rho$-integrals and summing contributions from the two patches,
we have the following results:
\begin{align}
S_\phi & = (t_{\rm max} - t_{\rm min}) |J|\,, \\
S_A 
&= 2 \times {2N \over \pi}
\int_{t_{\rm min}}^{t_{\rm max}}\!\! 
dt \int_{\rho_{\rm min}}^{\rho_{\rm max}}\!\! d\rho\, C_2
{
\cosh^4 \rho - C_1^2
\over
\cosh^2 \rho \sqrt{\cosh^2 \rho -C_1^2 - C_2^2 \coth^2 \rho}
} 
\notag \\
&= {4 N \over \pi} (t_{\rm max}-t_{\rm min}) 
C_2
{ 
\sqrt{
-C_1^2 \sinh^2 \rho_{\rm max}
+
\cosh^2 \rho_{\rm max}( \sinh^2 \rho_{\rm max} - C_2^2 )
}
\over 
\cosh \rho_{\rm max}
}\,. \notag 
\end{align}

\medskip 

Then let us consider \eqref{S_u}. It is composed of two terms. 
The first term gives $\alpha$- and $\ell$-independent contribution as 
explained in Appendix \ref{rho_int}, while 
the second term depends on $\alpha$ and $\ell$\,.  
By using $u=1/Z$ and substituting the solution, 
the second term can be rewritten as
\begin{equation}
2 \pi L^4 C_2 T_{\rm D3}
\sqrt{\cosh^2 \rho_{\rm max} - C_1^2 - C_2^2 \coth^2 \rho_{\rm max}}
\int_0^\pi \!d \varphi_1\,
\sin \varphi_1
\int_{t_{\rm min}}^{t_{\rm max}}\!\! dt\,
{Z' \over Z}\bigg|_{\rho_{\rm max}}\,.
\label{S_Z}
\end{equation}
Here $Z'/Z$ is given by
\begin{equation}
 {Z' \over Z} = - { A \cosh t + B \over C \cosh t + D}\,, \nonumber 
\end{equation} 
with the symbols $A$\,, $B$\,, $C$ and $D$ defined, respectively, as
\begin{align}
A &= \sinh \rho\,, \nonumber  \\ 
B &= 
\sqrt{1-\alpha^2} \cos \varphi_1 
( \cosh \rho \sin \chi + \sinh \rho \cos \chi \chi' )
+
\alpha ( \cosh \rho \cos \chi - \sinh \rho \sin \chi \chi' )\,, \nonumber \\
C & = \cosh \rho\,, \nonumber  \\
D & = \sinh \rho (\sqrt{1-\alpha^2} \sin \chi \cos \varphi_1 
+ \alpha \cos \chi )\,, \nonumber 
\end{align}
for the patch with $0 \leq \chi \leq \pi/2$. 
The terms for the other patch $\pi/2 \leq \chi \leq \pi$
are given by the usual replacement $\chi(\rho) \to \pi - \chi(\rho)$\,. 

\medskip 

The $t$-integral can analytically be performed and the result is  
\begin{equation}
 \int_{t_{\rm min}}^{t_{\rm max}}\!\! dt\, {Z' \over Z}
 =
 \bigg[
  -{A \over C} t
  - {2 ( -BC + AD ) \over C \sqrt{C^2-D^2} }
  \arctan 
  \bigg(
   - \sqrt{C-D \over C+D} \tanh \Big( { t \over 2 } \Big) 
  \bigg)
 \bigg]_{t_{\rm min}}^{t_{\rm max}}\,. \label{intZ'/Z}
\end{equation}
Then the $\varphi_1$-integral has to be performed. The first term in 
\eqref{intZ'/Z} does not depend on $\varphi_1$\,. 
Hence the $\varphi_1$-integral can be easily carried out and the result is 
\begin{equation}
- 2 \times {2 N \over \pi} C_2 (t_{\rm max} - t_{\rm min})
{ \sqrt{-C_1^2 \sinh^2 \rho_{\rm max} + \cosh^2 \rho_{\rm max} 
(\sinh^2 \rho_{\rm max} - C_2^2)}\over \cosh \rho_{\rm max} }\,. 
\nonumber 
\end{equation}
The overall factor $2$ comes from taking the 
two patches. This term exactly cancels the boundary term $S_A$\,.

\medskip

The $\varphi_1$-dependence of the second term in \eqref{intZ'/Z} is a little bit complicated. The $\varphi_1$-integral is evaluated in the large $|t_{\rm min,max}|$ and large $\rho_{\rm max}$ limit in Appendix
\ref{t_int}. The result of $\varphi_1$-integral depends on the patches
and also on $\alpha$\,.  The convergence of the integral
again assures that the result does not depend on 
$\ell$, i.e., possible dependences arise only through the cutoffs  
since $Z'/Z$ is independent of $\ell$\,. For the detailed calculation, 
see Appendix \ref{t_int}. 

\medskip 

After all, we have shown that  
\begin{equation}
  S_u  + S_A= {\rm const.} + 
\begin{cases}
 4 N C_2 \sqrt{1+C_2^2} & (\alpha=1)\\
0 & (\alpha \neq 1)
\end{cases}
\,. \nonumber 
\end{equation}

\medskip 

By gathering the results,  
$S_\phi+S_A+S_u$ has been evaluated as follows:
\begin{equation}
S_\phi+S_A+S_u = {\rm const.} +
2 |J| \log \bigg( { 2\ell \over \epsilon } \bigg)
+ 
\begin{cases}
 4 N C_2 \sqrt{1+C_2^2} & (\alpha=1) \\
0 & (\alpha \neq 1)
\end{cases}
\,. 
\label{res_S_boundary}
\end{equation}
The cutoff $\epsilon$ is defined as 
\[
- t_{\rm min} = t_{\rm max} \equiv \log (2 \ell / \epsilon)\,.
\] 
This definition is equivalent 
to the standard cutoff $Z=\epsilon$ imposed at the 
center $\rho=0$ of the D3-brane in the equal $t$ slice 
in terms of the original global coordinates. 

\subsection{Total action}

By gathering \eqref{res_SDBI_SWZ} and \eqref{res_S_boundary}, 
the total action is given by
\begin{align}
S_{\rm total} = f(C_1,C_2) + 2 |J| \log \Big( {2 \ell \over \epsilon} \Big) +
\begin{cases}
2N ( {\rm arcsinh}\,C_2 + C_2 \sqrt{1 + C_2^2} ) & ( \alpha=1) \\
0 & (\alpha \neq 1)
\end{cases} 
\,. \label{S[a,l]}
\end{align}
Here $f(C_1,C_2)$ is a function of $C_1$ and $C_2$\,. 
This reproduces the result of \cite{Miwa:2006vd} by setting $k=1$ and
taking large $N$\,. 

\medskip 

At first sight, it might be curious to find 
the result of Drukker-Fiol for a circular loop \cite{Drukker:2005kx}
with inverse sign in the third term of a straight line
($\alpha=1$)\,. But note that the third term
itself does not make sense, because it is just a part
of the total action.
Indeed, no one knows the normalization constant of the
gauge-theory operator and hence
even the overall normalization of $S_{\rm total}$ does not 
have any physical significance in our analysis. 
Remember that this is the case even in the string case
\cite{Miwa:2006vd}, where the authors pointed out also a subtlety
concerning a regularization in the presence of an R-charge.

\medskip 

A possible resolution proposed in
\cite{Miwa:2006vd} is to take the difference of the total action as
\begin{equation}
S_{\rm total}|_{\alpha \neq 1} - S_{\rm total}|_{\alpha=1}\,,\label{diff}
\end{equation}
and to compare it with the difference between a circular 
and a straight-line Wilson loop without R-charge.
In fact, the difference \eqref{diff} for our total action \eqref{S[a,l]}
properly reproduces the result of the Drukker-Fiol.
Inversely speaking, this prescription works well 
even for the D3-brane solution. Thus our result gives a non-trivial
support for the proposal in \cite{Miwa:2006vd}.

\subsubsection*{Consistency with $J\to 0$ limit} 

It may be interesting to see that \eqref{S[a,l]} is consistent with
$J\to 0$ limit. It is not obvious to check whether 
\eqref{S[a,l]} really reproduces the result of
\cite{Drukker:2005kx}.   
Here an ingredient of importance is the constant
term $f(C_1,C_2)$ in \eqref{S[a,l]}. 
In the case with $J=0$\,, we need to compute
$f(0,C_2)$\,.\footnote{More
precisely, it is necessary to evaluate (C.5) and (D.2) with
$C_1=0$\,. In particular, in computing (C.5), the only contribution
comes from the first term in (B.10) at $t=t_{\rm max}$\,.} 
It is still too complicated to do
analytically, so we have numerically evaluated it.
The result supports that 
\begin{eqnarray}
f(0,C_2) = -2N ({\rm arcsinh}\,C_2 +C_2{\textstyle \sqrt{1+C_2^2}})\,.
\end{eqnarray}
Thus \eqref{S[a,l]} reduces to 
\begin{align}
S_{\rm total} = \begin{cases}
0 & ( \alpha=1) \\
-2N ( {\rm arcsinh}\,C_2 + C_2 \sqrt{1 + C_2^2} )
& (\alpha \neq 1)
\end{cases} 
\,. \label{DFr}
\end{align}
This is nothing but the result of \cite{Drukker:2005kx}.
The mechanism to reproduce \eqref{DFr} is somewhat non-trivial, because 
it reappears from different integrals. 
The problem for the normalization of the Wilson loop might be 
clarified by investigating the behavior of $f(C_1,C_2)$ more in detail.  

\subsubsection*{Interpretation of $\ell$ dependence}

The $\ell$-dependence of \eqref{S[a,l]} should come from the contraction
of local operators inserted in the loop.  Then they have to have
conformal dimension $J$ due to the agreement of R-charge. This
$\ell$-dependence is consistent with the expectation value of the Wilson
loop with the insertions of $Z^J$ and its complex
conjugate \cite{Drukker:2006xg}.

\medskip 

However, the identification of \cite{Drukker:2006xg} 
should be modified to realize the fact that 
the solution is composed of the dual giant Wilson loop and a dual giant
graviton rather than a BPS particle. 
A key observation is that the $\ell$-dependence is
also consistent with the propagator of dual giant gravitons.
We will propose
another candidate of the dual gauge-theory operator for the D3-brane
solution in the next section.

\section{What is the corresponding  Wilson loop?}
\label{Wilson_Loop}

Finally let us discuss the Wilson loop corresponding to 
the D3-brane solution. The classical action computed in
the previous section should be an important key to identify it. 

\medskip

The local operator inserted in the loop should be modified by taking
account of the non-planar contributions. It is reasonable to consider a
dual giant graviton operator $Z_M{}^N$ as an inserted operator.  Here
$M$ and $N$ are the indices of $J$-th symmetric representation and its
conjugate representation.

\medskip

Thus a plausible candidate for the gauge-theory operator corresponding 
to the Euclidean D3-brane solution would possibly be the following:
\begin{equation} 
 \Gamma_{BN}^{CM} \tilde \Gamma_{DQ}^{AP}
 \big[ {\cal W}_{\vec X_f}^{\vec X_i} (C) \big]_A{}^B 
 \big[Z \big( \vec X_i \big) \big]_M{}^N 
 \big[ {\cal W}_{\vec X_i}^{\vec X_f} (C) \big]_C{}^D 
 \big[ \overline Z \big( \vec X_f \big) \big]_P{}^Q\,.
 \label{proposal}
\end{equation}
Here $\big[{\cal W}_{\vec X_i}^{\vec X_f} (C)\big]_A{}^B$ represents 
the $k$-th symmetric Wilson line running from 
$\vec X_i$ to $\vec X_f$ along the loop $C$\,. 
At $\vec X = \vec X_i $ and $\vec X_f$\,, 
there are four indices $(B, C, N, M)$ and $(D,A,Q,P)$\,, respectively. 
For the gauge invariance, these indices must be contracted separately 
at each of the points with some coefficients 
$\Gamma_{BN}^{CM}$ and $\widetilde \Gamma_{DQ}^{AP}$\,. 

\medskip

A simple way to contract the indices may be
taking $\Gamma_{BN}^{CM} = \delta_B{}^C \delta_N{}^M$ 
and $\widetilde \Gamma_{DQ}^{AP} = \delta_D{}^A \delta_Q{}^P$\,. 
Then the operator \eqref{proposal} just reduces to:
\begin{equation}
 {\rm Tr}_{{\rm S}_k} {\cal W}(C) \,
 {\rm Tr}_{{\rm S}_J} Z \big( \vec X_i \big) \,
 {\rm Tr}_{{\rm S}_J} \overline Z \big( \vec X_f \big)\,. 
\end{equation}
This is just the multiplication of Wilson loop without
local operator insertions with standard dual giant graviton operators,
and it does not reduce to \eqref{WZZ_MY}
when $k=1$ and $J \ll N$.

\medskip

An example of the operator which reduces to \eqref{WZZ_MY}
can be constructed by combining $k$-th and $J$-th symmetric indices 
into a $(k+J)$-th symmetric one.
We explain this type of operator by expressing 
the symmetric indices in terms of fundamental indices as: 
\begin{align}
& Z_M{}^N 
=
Z_{\{ m_1,\ldots,m_J \}}^{\{ n_1,\ldots,n_J \}}
\equiv
{\rm S}_{m_1,\ldots,m_J}^{\overline m_1,\ldots,\overline m_J}
{\rm S}_{\overline n_1,\ldots,\overline n_J}^{n_1,\ldots,n_J}
Z_{\overline m_1}^{\overline n_1}
\cdots
Z_{\overline m_J}^{\overline n_J}\,, \notag\\
& {\cal W}_A{}^B 
=
{\cal W}_{\{ a_1,\ldots, a_k \}}^{\{ b_1,\ldots, b_k \}}
\equiv
{\rm S}_{a_1,\ldots,a_k}^{\overline a_1,\ldots,\overline a_k}
{\rm S}_{\overline b_1,\ldots,\overline b_k}^{b_1,\ldots,b_k}
{\cal W}_{\overline a_1}^{\overline b_1}
\cdots
{\cal W}_{\overline a_k}^{\overline b_k}\,. \notag
\end{align}
All the lower-case indices express the (anti-) fundamental indices and the tensor 
${\rm S}_{m_1,\ldots,m_J}^{n_1,\ldots,n_J}$ is totally symmetric 
with respect to the upper (or lower) indices.
In this notation, the operator in which $k$-th and $J$-th symmetric indices 
are combined to $(k+J)$-th symmetric indices can be written down as
\begin{equation}
{\rm S}_{b_1,\ldots,b_k,n_1,\ldots,n_J}^{c_1,\ldots,c_k,m_1,\ldots,m_J}\, 
{\rm S}_{d_1,\ldots,d_k,q_1,\ldots,q_J}^{a_1,\ldots,a_k,p_1,\ldots,p_J}\,
 \big[ {\cal W}_{\vec X_f}^{\vec X_i}\big]_{\{a_1,\ldots,a_k\}}^{\{b_1,\ldots,b_k\}} 
 \big[Z \big( \vec X_i \big) \big]_{\{ m_1,\ldots,m_J \}}^{\{ n_1,\ldots,n_J\}} 
 \big[ {\cal W}_{\vec X_f}^{\vec X_i}\big]_{\{c_1,\ldots,c_k\}}^{\{d_1,\ldots,d_k\}} 
 \big[\overline Z \big( \vec X_f \big) \big]_{\{ p_1,\ldots,p_J \}}^{\{ q_1,\ldots,q_J\}}\,. 
 \notag
\end{equation}
In the case with $k=1$\,, this operator can be written as
(up to a normalization constant)
\begin{equation}
{\rm Tr}_{\rm F}
\Big[  
{\cal W}_{\vec X_f}^{\vec X_i} (C) Z(X_i)^J {\cal W}_{\vec X_i}^{\vec X_f} (C) 
\overline Z(\vec X_f)^J 
\Big]
+ 
\textrm{multi-trace operators}\,,
\end{equation}
and it reduces to the operator \eqref{WZZ_MY} 
when we assume $J \ll N$\,, since the multi-trace operators 
become sub-leading in $1/N$\,. 

\medskip

By assuming other coefficients $\Gamma_{BN}^{CM}$
and $\widetilde \Gamma_{DQ}^{AP}$, we can consider more generic 
Wilson loops with local operator insertions 
which seem to be consistent with 
the expectation value predicted by the D3-brane action. 
It would be nice to seek the definite 
choice of the coefficients $\Gamma_{BN}^{CM}$ and 
$\widetilde \Gamma_{DQ}^{AP}$\,, for example, via perturbative computation 
in the gauge-theory side. 
We leave this issue as a future work.

\section{Conclusion and discussion}
\label{Conclusion}

We have reexamined a rotating D3-brane solution 
in Lorentzian signature \cite{Drukker:2006zk} 
and discussed its tunneling picture. 

\medskip 

We first observed that the solution is composed of a Drukker-Fiol solution
\cite{Drukker:2005kx} and a dual giant graviton \cite{dGG1,dGG2}.
From this observation, we argued that the corresponding operator is a 
$k$-th symmetric Wilson loop with dual giant graviton operator insertions.

\medskip 

Then we have performed a double Wick rotation for the solution by
following the prescription of \cite{Dobashi:2002ar} and constructed the
solution in Euclidean AdS. For this solution, the total classical
action including appropriate boundary terms has been evaluated. The
resulting action reproduces the expectation value of the $k$-th
symmetric Wilson loop without local operator insertions
and the logarithmically divergent term which is
consistent with the correlation function of the dual giant graviton operators. 

\medskip 

These results may suggest that the usual prescription to compute 
the expectation value of Wilson loop by using the Gaussian matrix model 
can also be applied to the present case. 
It would be nice to find further supports in confirmation of 
the dual operator \eqref{proposal}. 
It is also necessary to have an argument to fix the coefficients 
of \eqref{proposal}. We leave these issues as future works.

\medskip

It would also be interesting to try to construct a rotating D5-brane. In
the case of giant Wilson loop the shape of D5-brane is
AdS$_2\times$S$^4$\,. Hence the S$^4$ part is expanding in S$^5$ and so
it seems difficult to find an appropriate ansatz in the same way as the case 
of dual giant Wilson loop. However, from our observation given in this
paper, we can easily guess that the desired solution should be composed
of an AdS$_2\times$S$^4$ D5-brane (giant Wilson loop) and a giant
graviton. By considering a giant spike solution \cite{GS} and deforming
it, it may be possible to find a rotating giant Wilson loop. The
corresponding Wilson loop in the gauge-theory side should be
obvious. All we have to do is to replace the $k$-th symmetric
representation and the dual giant graviton operator with the $k$-th
anti-symmetric representation and the giant graviton operator.

\medskip 

Furthermore it may be possible to construct a solution composed of dual
giant Wilson loop and giant graviton, or of giant Wilson loop and dual
giant graviton. It is nice to try to find such a solution.

\medskip 

It is also nice to study quantum fluctuations around the rotating D3-brane
solution. The fluctuations around the string solution of
\cite{Miwa:2006vd} have been discussed in \cite{SY1}.  The resulting
action is very complicated. But we can clearly see the asymptotic
behavior of the Lagrangian around the boundary and at the center of
AdS. It behaves as the semiclassical action around an AdS$_2$ solution
\cite{DGT}\footnote{For semiclassical approximation of DBI actions
around AdS-branes see \cite{SY2}.} around the boundary, while as the
pp-wave string at the center of AdS. The similar behavior should be
expected even for the fluctuations around the D3-brane solution.

\medskip 

We hope that the D-brane dynamics discussed in this paper would be an
important key to clarify some dynamical aspects of (dual) giant Wilson
loops.

\medskip 

\section*{Acknowledgment}

The authors would like to thank T.~Azeyanagi, K.~Hashimoto, S.~Iso, Y.~Kazama,
Y.~Kimura, Y.~Mitsuka, K.~Murakami, T.~Okuda, H.~Shimada, R.~Suzuki, 
D.~Trancanelli, A.~Tsuji, N.~Yokoi and T.~Yoneya for useful discussion. 
They also thank the Yukawa Institute for Theoretical Physics at Kyoto
University. Discussions during the YITP workshop YITP-W-07-05 on
``String Theory and Quantum Field Theory'' were useful to complete this
work.

\medskip

\noindent The work of A.~M.\ was supported in part by JSPS Research
Fellowships for Young Scientists. The work of K.~Y.\ was supported in
part by JSPS Postdoctoral Fellowships for Research Abroad and the
National Science Foundation under Grant No.\,NSF PHY05-51164.

\appendix

\section*{Appendix}

Hereafter we omit the subscript ``E'' of $t_{\rm E}$ and set $L=1$ for simplicity.

\section{Coordinate transformation} 
\label{transformation}

\medskip

The following decomposition of the coordinate transformation 
\eqref{trans_2} will be used in the next appendix:

\begin{enumerate}
\item Change from global coordinates to the Poincar\'e coordinates:
\begin{align}
&z = {e^t \over \cosh \rho}\,, \quad
r = e^t \tanh \rho\,, \label{1_1} \\
&x_1 = r \sin \chi \sin \varphi_1 \cos \varphi_2\,, \quad\! 
x_2 = r \sin \chi \sin \varphi_1 \sin \varphi_2\,, \quad\! \\
& x_3 = r \sin \chi \cos \varphi_1\,, \quad\!
x_4 = r \cos \chi\,. \label{1_2}
\end{align}
\item Rotation in the $(x_3,x_4)$-space:
\begin{align}
\bigg(
\begin{array}{c}
x_3'\\
x_4'
\end{array}
\bigg)
=
\bigg(
\begin{array}{cc}
\alpha & -\sqrt{1-\alpha^2} \\
\sqrt{1-\alpha^2} & \alpha
\end{array}
\bigg)
\bigg(
\begin{array}{c}
x_3 \\
x_4
\end{array}
\bigg)\,, \quad
x_{1,2}' = x_{1,2}\,, \quad
z'=z\,. \label{2}
\end{align}
\item Translation into the $x_4'$-direction:
\begin{equation}
x_4''=x_4'+1\,,\quad
x_{1,2,3}''=x_{1,2,3}'\,,\quad
z''=z'\,. \label{3}
\end{equation}

\item Inversion transformation and sign flip of $x_4''$:
\begin{equation}
Z''={z'' \over (x_i'')^2 + {z''}^2 }\,, \quad
X_{1,2,3}''=
{x_{1,2,3}'' \over (x_i'')^2+{z''}^2 }\,, \quad
X_4''= - {x_4'' \over (x_i'')^2 + {z''}^2 }\,. \label{4}
\end{equation}

\item Translation into the $X_4''$-direction:
\begin{equation}
X_4'=X_4''+{1 \over 2}\,, \quad
X_{1,2,3}'=X_{1,2,3}''\,, \quad
Z'=Z''\,. \label{5}
\end{equation}

\item Scale transformation in five dimensions:
\begin{equation}
Z = 2 \ell Z'\,, \quad X_i = 2\ell X_i'\,. \label{6}
\end{equation}
\end{enumerate}

\section{Derivation of the boundary three form
 $\Lambda_3^\alpha$}\label{C4_L3}

We shall derive the explicit form of $\Lambda_3^\alpha$ in
\eqref{P-P}. For this purpose it is convenient to consider ${\cal C}_4$
in \eqref{RR} as the $\alpha$-dependent four-form on the space spanned
by $(t,\rho,\chi,\varphi_1,\varphi_2)$ by using 
\eqref{trans_2}: 
\begin{equation}
{\cal C}_4[Z,X_i] = {\cal
 C}_4^\alpha[t,\rho,\chi,\varphi_1,\varphi_2]\,. 
\nonumber 
\end{equation}
Let us consider $\Lambda_4^\alpha$ defined as 
\begin{equation}
\Lambda_4^\alpha [t,\rho,\chi,\varphi_2,\varphi_2]
\equiv
{\cal C}_4^\alpha[t,\rho,\chi,\varphi_1,\varphi_2] 
- 
\widetilde C_4[t,\rho,\chi,\varphi_1,\varphi_2]\,. \nonumber 
\end{equation}
Here $\widetilde C_4[t,\rho,\chi,\varphi_1,\varphi_2]$ is defined as
\begin{equation}
\widetilde C_4[t,\rho,\chi,\varphi_1,\varphi_2] = 
 \sinh^4 \rho \sin^2 \chi \sin \varphi_1
 \bigg( dt + {d \rho \over \sinh \rho \cosh \rho} \bigg)
 \wedge d \chi \wedge d \varphi_1 \wedge d \varphi_2.
\end{equation}
By setting $\rho=\rho(\chi)$, it reduces to \eqref{4.7}.
With $\Lambda_4^\alpha$\,, $d \Lambda_3^\alpha$ in
\eqref{P-P} is rewritten as
\begin{equation}
d \Lambda_3^\alpha 
= \Lambda_4^\alpha[t,\rho=\rho(\chi),\chi,\varphi_1,\varphi_2]\,.\nonumber
\end{equation}
First it is easy to check that $\widetilde C_4[t,\rho,\chi,\varphi_1,\varphi_2]$
is rewritten as 
\begin{equation}
 \widetilde C_4[t,\rho,\chi,\varphi_1,\varphi_2]
 = { 1 \over z^4 } dx_4 \wedge dx_1 \wedge dx_2 \wedge dx_3\,.  
 \label{app_C_4}
\end{equation}
Here $(z,x_i)$ are related to $(t,\rho,\chi,\varphi_1,\varphi_2)$ via
\eqref{1_1}--\eqref{1_2}.
Since the steps 2., 3., 5., and 6. in the previous subsection 
keep the form of the four form potentials \eqref{RR} and \eqref{app_C_4}, 
we have 
\begin{align}
\Lambda_4^\alpha
&\equiv
{1 \over Z^4} d X_4 \wedge d X_1 \wedge d X_2 \wedge d X_3
-
{1 \over z^4} d x_4 \wedge d x_1 \wedge d x_2 \wedge d x_3 \notag \\
&=
{1 \over {Z''}^4} d X''_4 \wedge d X''_1 \wedge d X''_2 \wedge d X''_3
-
{1 \over {z''}^4} d x''_4 \wedge d x''_1 \wedge d x''_2 \wedge d x''_3\,. 
\notag
\end{align}
Here $(Z'',X_i'')$ and $(z'',x_i'')$ are related to
$(t,\rho,\chi,\varphi_1,\varphi_2)$ via \eqref{1_1}--\eqref{4}. 
With help of \eqref{4} we can
write $\Lambda_4^\alpha$ in terms of $(z'',x_i'')$ as:
\begin{align}
 &\Lambda_4^\alpha = {2 \over {z''}^4 } {1 \over g} 
 ( z'' d z'' \wedge \Delta_3 - {z''}^2 \Omega_4)\,, \label{dLambda_3} \\
 & \Delta_3 
 = (x_4'' dx_1'' - x_1'' dx_4'') \wedge dx_2'' \wedge dx_3''
   +
   (x_2'' dx_3'' - x_3'' dx_2'') \wedge dx_4'' \wedge dx_1''\,,\notag \\
 & \Omega_4 
 = d x_4'' \wedge d x_1'' \wedge d x_2'' \wedge d x_3''\,,\qquad 
g = {z''}^2 + (x_i'')^2\,. \notag
\end{align}
We further introduce the polar coordinates 
$(r'',\theta_1,\theta_2,\theta_3)$ defined as 
\begin{eqnarray}
&&  x_1'' = r'' \sin \theta_1 \sin \theta_2 \cos \theta_3\,, \quad
 x_2'' = r'' \sin \theta_1 \sin \theta_2 \sin \theta_3\,, \nonumber \\ 
&& x_3'' = r'' \sin \theta_1 \cos \theta_2\,, \quad
 x_4'' = r'' \cos \theta_1\,.  \nonumber 
\end{eqnarray}
By using them, $\Delta_3$ and $\Omega_4$ can be rewritten into the
following forms: 
\begin{align}
\Delta_3 
& = 
{r''}^4 \sin^2 \theta_1 \sin \theta_2 
d \theta_1 \wedge d \theta_2 \wedge d \theta_3\,, \notag \\
\Omega_4 
& =
{r''}^3 \sin^2 \theta_1 \sin \theta_2 
dr'' \wedge d \theta_1 \wedge d \theta_2 \wedge d \theta_3\,. \notag
\end{align}
Then $\Lambda_4^\alpha$ is also rewritten as the exact form:
\begin{equation}
\Lambda_4^\alpha = d \widetilde \Lambda_3^\alpha\,, \qquad
\widetilde \Lambda_3^\alpha = 
\bigg( - { {r''}^2 \over {z''}^2} + 
\log\bigg(1 + { {r''}^2 \over {z''}^2}\bigg)\bigg)
{\Delta_3 \over {r''}^4}\,. \label{Lambda3''}
\end{equation}
With \eqref{1_1}--\eqref{3}\,, $\tilde \Lambda_3^\alpha$ and $\Delta_3$
can be expressed in terms of $(t,\rho,\chi,\varphi_1,\varphi_2)$\,. 
Then $z''$ and $r''$ can be rewritten as
\begin{equation}
{z''}^2 = {e^{2t} \over \cosh^2 \rho}\,, \quad {r''}^2 = 
e^{2t} \tanh^2 \rho 
+ 
2 e^t \tanh \rho (\alpha \cos \chi 
+ 
\sqrt{1-\alpha^2} \sin\chi \cos \varphi_1) 
+ 
1\,. \label{2z''2r''}
\end{equation}
Now $\Delta_3$ is given by 
\begin{align}
\Delta_3
&=
(\Delta_3)_{t\chi\varphi_2}
dt \wedge d\chi \wedge d\varphi_2 
+
(\Delta_3)_{\rho \chi \varphi_2} 
d\rho \wedge d \chi \wedge d \varphi_2
+
(\Delta_3)_{t \varphi_1 \varphi_2} 
dt \wedge d \varphi_1 \wedge d\varphi_2 \notag\\
&\qquad+
(\Delta_3)_{\rho \varphi_1 \varphi_2}
d\rho \wedge d \varphi_1 \wedge d \varphi_2
+
(\Delta_3)_{\chi \varphi_1 \varphi_2}
d \chi \wedge d \varphi_1 \wedge d \varphi_2\,, \notag 
\end{align}
where the non-vanishing components are
\begin{align}
(\Delta_3)_{t\chi\varphi_2}
& =
-\sqrt{1-\alpha^2}e^{3t} 
\tanh^3 \rho \sin \chi \sin^2 \varphi_1\,, \label{D3_1}\\
(\Delta_3)_{\rho \chi \varphi_2} 
& =
-\sqrt{1-\alpha^2}e^{3t} 
{\tanh^2 \rho \over \cosh^2 \rho} \sin \chi \sin^2 \varphi_1\,, 
\label{D3_2}  \\
(\Delta_3)_{t \varphi_1 \varphi_2} 
& =
\alpha e^{3t} \tanh^3 \rho \sin^3 \chi \sin \varphi_1
-
\sqrt{1-\alpha^2} e^{3t} 
\tanh^3 \rho \sin^2 \chi \cos \chi \sin \varphi_1 \cos \varphi_1
\,, \label{D3_3} \\
(\Delta_3)_{\rho \varphi_1 \varphi_2}
& =
\alpha e^{3t} { \tanh^2 \rho \over \cosh^2 \rho}
\sin^3 \chi \sin \varphi_1
-
\sqrt{1-\alpha^2}e^{3t}
{\tanh^2 \rho \over \cosh^2 \rho} 
\sin^2 \chi \cos \chi \sin \varphi_1 \cos \varphi_1
\,, \label{D3_4}\\
(\Delta_3)_{\chi \varphi_1 \varphi_2}
& =
e^{4t} \tanh^4 \rho \sin^2 \chi \sin \varphi_1+
\alpha e^{3t} \tanh^3 \rho \sin^2 \chi \cos \chi \sin \varphi_1  
\notag \label{D3_5}\\
& \hspace{6cm}
+ \sqrt{1-\alpha^2} e^{3t} \tanh^3 \rho \sin^3 \chi \sin \varphi_1 
\cos \varphi_1\,.
\end{align}
Finally $\Lambda_3^\alpha$ is given by
\begin{equation}
\Lambda_3^\alpha [t,\chi,\varphi_1,\varphi_2] 
= 
\widetilde \Lambda_3^\alpha [t,\rho=\rho(\chi),\chi,\varphi_1,\varphi_2]\,.
\label{LtL}
\end{equation}

\section{Integral of $(\Lambda_3^\alpha)_{t \varphi_1 \varphi_2}$ and 
$(\Lambda_3^\alpha)_{\chi \varphi_1 \varphi_2}$}\label{intL3}

\subsection{$(\Lambda_3^\alpha)_{t \varphi_1 \varphi_2}$}
\label{app_intL3_tvv}

The aim here is to perform the integral, 
\begin{equation}
\int_{t_{\rm min}}^{t_{\rm max}}\!\!\! dt
\int_0^\pi \!\! d \varphi_1 
\int_0^{2 \pi}\!\!\! d\varphi_2\,
\Big[
(\Lambda_3^\alpha)_{t \varphi_1 \varphi_2} 
\Big]_{\chi_{\rm min}}^{\chi_{\rm max}}\,.
\label{app_int_L3}
\end{equation}
The explicit form of the integrand is given by \eqref{LtL} with
\eqref{Lambda3''} and \eqref{D3_3}.  For $\alpha \neq 1$\,, this
integral vanishes in the limit $\rho_{\rm max} \to \infty$ or
equivalently in the limit $\chi_{\rm min} \to 0$ and $\chi_{\rm
max}\to\pi$\,.  This can be shown by rewriting the integral
\eqref{app_int_L3} in the form in which a single $\sin \chi$ is
extracted as an overall factor, i.e., in the form as $(\sin \chi) \times
\int dt (\cdots)$\,. In this form we can show that the $t$-integral
still converges.  Since the extracted overall factor $\sin \chi$
vanishes in the limit $\rho_{\rm max} \to \infty$\,, \eqref{app_int_L3}
vanishes.  

\medskip

However the case with $\alpha=1$ is special and then 
$r''$ at $t=0$ is given by 
\begin{equation}
{r''}^2 = 
\tanh^2 \rho 
+ 
2 \tanh \rho \cos \chi 
+ 
1 \qquad \textrm{(at $t=0$)}\,. \label{c2} 
\end{equation}
Therefore $r''$ becomes zero at the upper boundary $\chi=\chi_{\rm max}$
taking the limit $\chi_{\rm max} \to \pi$\,. 
We can also check that ${r''}^2/{z''}^2 \to C_2^2$ in the same limit. 
This means that, for finite $C_2$\,, the
integrand of \eqref{app_int_L3} tends to diverge at $t=0$ in the limit
$\chi_{\rm max} \to \pi$\,.
This divergence cancels the small overall
factor $\sin \chi$ and hence the integral \eqref{app_int_L3} may give a
finite value to 
$S_{\rm WZ}$\,. Nevertheless most terms in the integral actually
vanish apart from the following two integrals: 
\begin{enumerate}
\item 
\begin{align}
 &\int \!\! dt d \varphi_1 d \varphi_2\,
 \bigg[
 -{ 1 \over {z''}^2 {r''}^2} e^{3t} \tanh^3 \rho \sin^3 \chi \sin \varphi_1
 \bigg]_{\chi_{\rm min}}^{\chi_{\rm max}} \notag \\
&=
- 4 \pi \int_{t_{\rm min}}^{t_{\rm max}}\!\! dt\,
\bigg[
 {e^t \cosh^2 \rho  \tanh^3 \rho \sin^3 \chi
 \over
 e^{2t} \tanh^2 \rho + 2 e^t \tanh \rho \cos \chi + 1}
\bigg]_{\chi_{\rm min}}^{\chi_{\rm max}} \notag \\
&=
4 \pi
\bigg[\sinh^2 \rho \sin^2 \chi
\bigg(
\arctan \bigg[ {1 + \tanh \rho \cos \chi \over \tanh \rho \sin \chi} \bigg]
+\arctan \bigg[ {\tanh \rho + \cos \chi \over \sin \chi } \bigg] \notag \\
& \hspace{1.5cm}
- \arctan \bigg[ { \tanh \rho \cos \chi  
+ e^{-t_{\rm min}} \over \tanh \rho \sin \chi } \bigg]
- \arctan \bigg[ { \cos \chi 
+ e^{t_{\rm max}} \tanh \rho \over \sin \chi} \bigg]
\bigg)
\bigg]_{\chi_{\rm min}}^{\chi_{\rm max}} \notag \\
& \to 4 \pi C_2^2
\bigg[
\Big(0+0-{\pi \over 2} - {\pi \over 2} \Big)
-
\Big({\pi \over 2} + {\pi \over 2} - {\pi \over 2} - {\pi \over 2}\Big)
\bigg]\hspace{1cm} (|t_{\rm min,max}|\,, \,\rho_{\rm max} \to \infty) \label{--+} \\
& = -4 \pi^2 C_2^2\,. \notag 
\end{align}
In \eqref{--+}, among two contributions $-\pi$ and $0$ in the large
round bracket, the first one, i.e., $-\pi$\,, is the contribution from
$\chi=\chi_{\rm max}$ and the second vanishing term is from the lower
boundary $\chi=\chi_{\rm mix}$\,.
\item
\begin{align}
&\int\!\! dt d \varphi_1 d \varphi_2\,
 \bigg[
  {1 \over {r''}^4}
  \log\bigg( 1 + {{r''}^2 \over {z''}^2} \bigg)
  e^{3t} \tanh^3 \rho \sin^3 \chi \sin \varphi_1
 \bigg]_{\chi_{\rm min}}^{\chi_{\rm max}} \notag \\
& \hspace*{-0.5cm} = 
4\pi
\bigg[
 \tanh^3 \rho \sin^2 \chi \int\!\! dt \sin \chi e^t{ 
-t + 2 \log ( \cosh \rho )
+
\log
(
e^t + e^{-t} + 2 \tanh \rho 
\cos \chi
)
\over 
(e^t \tanh^2 \rho 
+ 
2 \tanh \rho \cos \chi 
+ 
e^{-t})^2} 
\bigg]_{\chi_{\rm min}}^{\chi_{\rm max}}\,.  \label{3-1}
\end{align}
On the lower boundary $\chi=\chi_{\rm min}$\,, the integrand develops no
singularity in the limit $\chi_{\rm min} \to 0$\,. As for the behavior of
the integrand at the boundary $t \sim t_{\rm min}\,, t_{\rm max}$ of the
domain of integration, we have
\begin{equation}
\textrm{the integrand of \eqref{3-1}} \to 
\begin{cases}
e^{-t_{\rm max}} 
{2 \log(\cosh \rho_{\rm max}) 
\over \tanh^4 \rho_{\rm max} } \sin \chi_{\rm min}\\
-2 t_{\rm min} e^{3t_{\rm min}} \sin \chi_{\rm min}
\end{cases}
\,. \nonumber 
\end{equation}
Hence the integral is finite in the large $\rho_{\rm max}$ and the large
$|t_{\rm min,max}|$ limit. On the other hand, the extra overall factor
becomes zero in this limit:
\begin{equation}
 \tanh^3 \rho_{\rm max}\, \sin^{2} \chi_{\rm min} \to 0\,. \nonumber 
\end{equation}
Thus we have 
\begin{equation}
 \textrm{the lower boundary contribution of } \eqref{3-1} \to 0\,. \nonumber 
\end{equation}

At the upper boundary $\chi=\chi_{\rm max}$\,, by the numerical
analysis, we found that
\begin{equation}
 \textrm{the upper boundary contribution of } \eqref{3-1} \to 
 4 \pi^2 \big( C_2^2 - C_2 \sqrt{1+C_2^2} + {\rm arcsinh} C_2 \big)\,. 
\nonumber 
\end{equation}
In summary, we obtain that
\begin{equation}
 \eqref{3-1} \to 
 4 \pi^2 \big(- C_2 \sqrt{1+C_2^2} + {\rm arcsinh} C_2 \big)\,. 
\nonumber 
\end{equation}

\end{enumerate}

\subsection{$ (\Lambda_3^\alpha)_{\chi \varphi_1 \varphi_2}$}
\label{app_intL3_cvv}

The aim here is to evaluate the 
$\alpha$- and $\ell$-dependence of the integral:
\begin{equation}
\int_{\chi_{\rm min}}^{\chi_{\rm max}}\!\! d\chi
\int_0^\pi \!\! d \varphi_1
\int_0^{2 \pi} \!\!\! d\varphi_2\,
\Big[
(\Lambda_3^\alpha)_{\chi \varphi_1 \varphi_2}
\Big]_{t_{\rm min}}^{t_{\rm max}}\,. 
\label{int_L3_cvv}
\end{equation}
The integrand is given 
by \eqref{LtL} with \eqref{Lambda3''}, \eqref{D3_4} and \eqref{D3_5}.
Although the integrand depends only on $\alpha$\,,
$\ell$-dependence may arise through the cutoffs. 

\medskip

Let us examine each of the terms in the integrand. 
First, the integrals of the terms in 
\eqref{D3_4} and \eqref{D3_5} 
which are proportional to $e^{3t}$  
vanish in the limit $|t_{\rm min,max}|\to \infty$\,. 
This is because 
the integrals of these terms can be rewritten as $e^{-|t_{\rm min,max}|}
\times \int d\chi(\cdots)$ in which the $\chi$-integral gives no
divergence. 

\medskip 

Next let us consider the term linear in $e^{4t}$ in \eqref{D3_5} whose
power is greater than the previous case by $e^t$\,.
Although the contribution from the lower edge, $t=t_{\rm min}$\,, vanishes,
there may be non-trivial contribution from the upper edge, $t=t_{\rm max}$\,. 
It is easy to check the convergence of the integral in the limit
$t_{\rm max} \to \infty$ and $\rho_{\rm max} \to \infty$\,. 
What is more we can also check that the contribution does not depend on $\alpha$\,. 
This is essentially because the $\alpha$-dependence is sub-leading 
with respect to $e^{-t_{\rm max}}$\,, 
as can be seen from the expression of $r''$ in \eqref{2z''2r''}\,.\footnote{
In the case with $\rho_{\rm min}=0$\,,  
we need to consider $\rho$-integral instead of $\chi$-integral.
Then for the range $\rho \sim 0$ 
the same argument can not be applied since $r''$ can vanish.
However, the $\alpha$-independence can easily be checked even for 
the range by Taylor expanding the $\log$ term in \eqref{Lambda3''}. 
} 
Hence we understand that the integral \eqref{int_L3_cvv}
does not depend on $\alpha$ nor on $\ell$ 
in the large $|t_{\rm min,max}|$ and $\rho_{\rm max}$ limit.

\section{Evaluation of $\boldsymbol S_u$} \label{boundary}

\medskip

The boundary term $S_u$ is composed of the two terms like
\begin{equation}
S_u =
\int_{\rho_{\rm min}}^{\rho_{\rm max}}\!\! d\rho d\varphi_1 d\varphi_2\, 
\bigg[
{\partial {\cal L} \over \partial \dot Z } Z 
\bigg]_{t_{\rm min}}^{t_{\rm max}}
+
\int_{t_{\rm min}}^{t_{\rm max}}\!\! dt d\varphi_1 d\varphi_2\,
{\partial {\cal L} \over \partial Z'} Z\bigg|_{\rho_{\rm max}}\,. 
\label{app_S_u}
\end{equation}

\subsection{$\rho$-integral}
\label{rho_int}

It is easy to check that the large 
$|t_{\rm min,max}|$ and $\rho_{\rm max}$ limit 
of the first term in \eqref{app_S_u}
converges to give the following expression:
\begin{equation}
-2J - {8N \over \pi} C_2 \int d \rho 
{
(\cosh^2 \rho - C_1^2 )^2 + C_2^2 \cosh^4 \rho 
\over 
\cosh^2 \rho (\cosh^2 \rho - C_1^2) 
\sqrt{\cosh^2 \rho -C_1^2 -C_2^2 \coth^2 \rho}}.
\label{-2J-intdrho}
\end{equation}
Here we have summed contributions from two patches.
From this expression, it is clear that the first term of 
\eqref{app_S_u} does not depend on $\alpha$ and $\ell$\,.
In the case with $k \sim 1$ and $C_1=1$\,,
\eqref{-2J-intdrho} is reduced to the result of \cite{Miwa:2006vd} as
\begin{equation}
-2J - {2 k \sqrt{\lambda } \over \pi}.
\end{equation}

\subsection{$t$-integral}
\label{t_int}

The $t$-integral of \eqref{app_S_u} is given by \eqref{S_Z}--\eqref{intZ'/Z}.
In the main text, we have performed the $\varphi_1$-integral of the first term 
of \eqref{intZ'/Z}. We consider here the second term.  In the large $| t_{\rm
min,max} |$ and large $\rho_{\rm max}$ limit, the second term of
\eqref{intZ'/Z} will be estimated as
\begin{enumerate}
\item $\alpha=1$
\begin{align}
-BC+AD & \to 
\begin{cases}
 -1 - C_2^2 & (\chi=\chi_{\rm min}) \\
 1 + C_2^2  & (\chi=\chi_{\rm max})
\end{cases}
\,, \notag \\
 C \sqrt{C^2-D^2} & \to
\begin{cases}
\cosh \rho_{\rm max} \sqrt{1+C_2^2} & (\chi=\chi_{\rm min} ) \\
\cosh \rho_{\rm max} \sqrt{1+C_2^2} & (\chi=\chi_{\rm max} )
\end{cases}
\,,\notag \\
\sqrt{ C-D \over C+D } \tanh\Big({t\over2}\Big)
& \to
\begin{cases}
0 & (\chi=\chi_{\rm min}, \,\, t=t_{\max} ) \\
0 & (\chi=\chi_{\rm min}, \,\, t=t_{\min} ) \\
+\infty & (\chi=\chi_{\rm max}, \,\, t=t_{\max} ) \\
-\infty & (\chi=\chi_{\rm max}, \,\, t=t_{\max} )
\end{cases}
\,. \notag 
\end{align}

The integrand of $\varphi_1$-integral becomes as follows:
\begin{align}
\begin{cases}
0 &(\chi=\chi_{\rm min}, \,\, t=t_{\max} ) \\
0 & (\chi=\chi_{\rm min}, \,\, t=t_{\min} ) \\
N C_2 \sqrt{1+C_2^2} \sin \varphi_1 & (\chi=\chi_{\rm max}, \,\, 
t=t_{\max} ) \\
-N C_2 \sqrt{1+C_2^2} \sin \varphi_1 & (\chi=\chi_{\rm max}, \,\, 
t=t_{\min} )
\end{cases}
\,. \notag 
\end{align}

\item $\alpha\neq1$
\begin{align}
-BC+AD &\to 
\begin{cases}
C_2\sqrt{1-\alpha^2}\cos \varphi_1 \cosh \rho_{\rm max}
& (\chi=\chi_{\rm min} ) \\
C_2\sqrt{1-\alpha^2}\cos \varphi_1 \cosh \rho_{\rm max}
& (\chi=\chi_{\rm max} )
\end{cases}
\,, \notag \\
C \sqrt{C^2-D^2} & \to
\begin{cases}
 \sqrt{1-\alpha^2} \cosh^2 \rho_{\rm max}& (\chi=\chi_{\rm min} ) \\
 \sqrt{1-\alpha^2} \cosh^2 \rho_{\rm max}& (\chi=\chi_{\rm max} )
\end{cases}
\,, \notag \\
\sqrt{ C-D \over C+D } \tanh\Big({t\over2}\Big)
& \to
\begin{cases}
\displaystyle{\sqrt{1-\alpha \over 1 + \alpha}}& 
(\chi=\chi_{\rm min}, \,\, t=t_{\max} ) \\
-\displaystyle{\sqrt{1-\alpha \over 1 + \alpha}}& 
(\chi=\chi_{\rm min}, \,\, t=t_{\min} ) \\
\displaystyle{\sqrt{1+\alpha \over 1 - \alpha}}& 
(\chi=\chi_{\rm max}, \,\, t=t_{\max} ) \\
-\displaystyle{\sqrt{1+\alpha \over 1 - \alpha}}& 
(\chi=\chi_{\rm max}, \,\, t=t_{\min} )
\end{cases} 
\,. \notag
\end{align}
The integrand of $\varphi_1$-integral becomes as follows:
\begin{align}
\begin{cases}
(2 N / \pi) C_2^2 \sin \varphi_1 \cos \varphi_1
\displaystyle{\arctan \bigg(\sqrt{1-\alpha \over 1+ \alpha} \bigg)}
&  (\chi=\chi_{\rm min}, \,\, t=t_{\max} ) \\
-(2 N / \pi) C_2^2 \sin \varphi_1 \cos \varphi_1
\displaystyle{\arctan \bigg(\sqrt{1-\alpha \over 1+ \alpha} \bigg)}
& (\chi=\chi_{\rm min}, \,\, t=t_{\min}) \\
(2 N / \pi) C_2^2 \sin \varphi_1 \cos \varphi_1
\displaystyle{\arctan \bigg(\sqrt{1+\alpha \over 1 - \alpha} \bigg)}
& (\chi=\chi_{\rm max}, \,\, t=t_{\max}) \\
-(2 N / \pi) C_2^2 \sin \varphi_1 \cos \varphi_1
\displaystyle{\arctan \bigg(\sqrt{1+\alpha \over 1- \alpha} \bigg)}
& (\chi=\chi_{\rm max}, \,\, t=t_{\min})
\end{cases}
\,.\notag
\end{align}
\end{enumerate}

By performing $\varphi_1$-integral, we have 
\begin{enumerate}
\item $\alpha=1$:
\begin{align}
 \int_{t_{\rm min}}^{t_{\rm max}}\!\! dt d \varphi_1 d \varphi_2\, 
 { \partial {\cal L} \over \partial Z'} Z 
 \bigg|_{\chi_{\rm max}}\!\!\!\!\!\!
 =
 -{1 \over 2} S_A + 4 N C_2 \sqrt{1+C_2^2}\,, \quad\!
  \int_{t_{\rm min}}^{t_{\rm max}}\!\!dt d\varphi_1 d \varphi_2\, 
 { \partial {\cal L} \over \partial Z'} Z 
 \bigg|_{\chi_{\rm min}}\!\!\!\!\!\!
 =
 -{1 \over 2} S_A\,. \notag
\end{align}
\item $\alpha\neq1$\,:
\begin{align}
 \int_{t_{\rm min}}^{t_{\rm max}}\!\! dt d \varphi_1 d \varphi_2\, 
 { \partial {\cal L} \over \partial Z'} Z 
 \bigg|_{\chi_{\rm max}}
 =
  \int_{t_{\rm min}}^{t_{\rm max}} \!\!dt d \varphi_1 d \varphi_2\, 
 { \partial {\cal L} \over \partial Z'} Z 
 \bigg|_{\chi_{\rm min}}
 =
 -{1 \over 2} S_A\,. \notag
\end{align}
\end{enumerate}

\end{document}